\documentclass{article}

\usepackage{arxivsimplified}

\usepackage[utf8]{inputenc} 
\usepackage[T1]{fontenc}    
\usepackage{hyperref}       
\usepackage{url}            
\usepackage{booktabs}       
\usepackage{amsfonts}       
\usepackage{nicefrac}       
\usepackage{microtype}      
\usepackage{lipsum}

\usepackage{graphicx}
\graphicspath{{figs_pdf/}}
\usepackage{subfigure}
\usepackage{cite}
\usepackage[dvipsnames]{xcolor}
\usepackage{amsmath}
\usepackage{blindtext}
\usepackage{amsfonts}
\usepackage{multirow}
\usepackage{array}
\usepackage{mathtools}
\usepackage{listings}
\usepackage{xcolor}
\definecolor{codegreen}{rgb}{0,0.6,0}
\definecolor{codegray}{rgb}{0.5,0.5,0.5}
\definecolor{codepurple}{rgb}{0.58,0,0.82}
\definecolor{backcolour}{rgb}{0.95,0.95,0.92}
\usepackage{color}
\usepackage{xcolor}
\definecolor{rev1}{rgb}{0,0,0}
\definecolor{rev2}{rgb}{0,0,0}
\definecolor{rev3}{rgb}{0,0,0}
\usepackage{esvect}
\usepackage{tabularx}
\newcolumntype{L}[1]{>{\raggedright\arraybackslash}p{#1}}
\newcolumntype{C}[1]{>{\centering\arraybackslash}p{#1}}
\newcolumntype{R}[1]{>{\raggedleft\arraybackslash}p{#1}}

\usepackage{algorithm}
\usepackage{algorithmic}
\usepackage{enumitem}
\setlist[itemize]{leftmargin=*}

\lstdefinestyle{mystyle}{
    backgroundcolor=\color{backcolour},   
    commentstyle=\color{codegreen},
    keywordstyle=\color{magenta},
    numberstyle=\tiny\color{codegray},
    stringstyle=\color{codepurple},
    basicstyle=\ttfamily\footnotesize,
    breakatwhitespace=false,         
    breaklines=true,                 
    captionpos=b,                    
    keepspaces=true,                 
    numbers=left,                    
    numbersep=5pt,                  
    showspaces=false,                
    showstringspaces=false,
    showtabs=false,                  
    tabsize=2
}
 
\usepackage{scalerel,stackengine}
\stackMath
\newcommand\reallywidehat[1]{%
\savestack{\tmpbox}{\stretchto{%
  \scaleto{%
    \scalerel*[\widthof{\ensuremath{#1}}]{\kern-.6pt\bigwedge\kern-.6pt}%
    {\rule[-\textheight/2]{1ex}{\textheight}}
  }{\textheight}%
}{0.5ex}}%
\stackon[1pt]{#1}{\tmpbox}%
}

\title{A nonintrusive hybrid neural-physics modeling of incomplete dynamical systems: Lorenz equations}

\author{
    Suraj Pawar  \\
    School of Mechanical \& Aerospace Engineering,\\
    Oklahoma State University, \\
    Stillwater, Oklahoma - 74078, USA.\\
    \texttt{supawar@okstate.edu} \\
    \And
    Omer San \\
    School of Mechanical \& Aerospace Engineering,\\
    Oklahoma State University, \\
    Stillwater, Oklahoma - 74078, USA.\\
    \texttt{osan@okstate.edu} \\
    \And
    Adil Rasheed \\
	Department of Engineering Cybernetics, \\
	Norwegian University of Science and Technology, \\
	7465 Trondheim, Norway. \\
    Department of Mathematics and Cybernetics, \\
    SINTEF Digital, \\
    7034 Trondheim, Norway. \\
    \AND
    Ionel M. Navon  \\
    Department of Scientific Computing,\\
    Florida State University, \\
    Tallahassee Florida - 32306, USA
}

\begin{document}
\maketitle

\begin{abstract}
This work presents a hybrid modeling approach to data-driven learning and representation of unknown physical processes and closure parameterizations. These hybrid models are suitable for situations where the mechanistic description of dynamics of some variables is unknown, but reasonably accurate observational data can be obtained for the evolution of the state of the system. \textcolor{rev3}{In this work, we propose machine learning to account for missing physics and then data assimilation to correct the prediction. In particular, we devise an effective methodology based on a recurrent neural network to model the unknown dynamics. A long short-term memory (LSTM) based correction term is added to the predictive model in order to take into account hidden physics.} \textcolor{rev3}{Since LSTM introduces a black-box approach for the unknown part of the model, we investigate whether the proposed hybrid neural-physical model can be further corrected through a sequential data assimilation step.} We apply this framework to the weakly nonlinear Lorenz model that displays quasiperiodic oscillations, the highly nonlinear chaotic Lorenz model, \textcolor{rev1}{and two-scale Lorenz model}. The hybrid neural-physics model \textcolor{rev3}{yields accurate results for} the weakly nonlinear Lorenz model with the predicted state close to the true Lorenz model trajectory. For the highly nonlinear Lorenz model \textcolor{rev1}{and the two-scale Lorenz model}, the hybrid neural-physics model deviates from the true state due to the accumulation of prediction error from one time step to the next time step. \textcolor{rev3}{The ensemble Kalman filter approach takes into account the prediction error and updates the diverged prediction using available observations in order to provide a more accurate state estimate for the highly nonlinear chaotic Lorenz model and two-scale Lorenz system.} The successful synergistic integration of neural network and data assimilation for low-dimensional system shows the potential benefits of the proposed hybrid-neural physics model for complex dynamical systems.      
 
\end{abstract}


\section{Introduction} \label{sec:introduction}
The advancement in high-performance computing has made numerical simulation of atmosphere and climate models on computational grid consisting of $O(10^7)$ cells, spaced $O(10~$\text{km}$)$ - $O(100~$\text{km}$)$ possible \cite{schneider2017earth}. However, the geophysical flows are governed by scales smaller than the mesh size of the climate models, and resolving these scales is essential for accurate prediction. For example, the dynamical scales of boundary layer clouds are $O(10~\text{m})$ \cite{wood2012stratocumulus}, and the submesoscale dynamics of oceans have length scales of $O(100~\text{m})$ \cite{fox2014principles}, and will remain unresolvable in the near future \cite{schneider2017climate}. Therefore, the physics in climate models is essentially incomplete and the dynamics of smaller scales in the atmosphere and oceans must be represented through some parameterization schemes. Another example of an incomplete dynamical system is a hybrid coupled model where the physical equations are replaced by empirical relations to achieve a large computational speedup and greater stability. One example of a hybrid coupled model is the coupling of the statistical atmospheric model with a fully nonlinear ocean general circulation model, where the wind stress is empirically estimated from the ocean model variables \cite{barnett1993enso}. 

The dynamics of the missing physics in geophysical flows is inherently nonlinear and replacing it with linear statistical methods will not deliver accurate prediction for longer periods. The `universal approximator theorem' \cite{hornik1991approximation} of the neural network makes them an attractive choice for learning nonlinear dynamics in incomplete dynamical systems. Some of the early use of the neural network for dynamical systems are turbulence control for drag reduction \cite{lee1997application}, prediction of wind stress field from the ocean state in a hybrid coupled atmosphere-ocean model \cite{tang2001neural}, modeling missing dynamics of the Lorenz system \cite{tang2001coupling,liaqat2003applying}, and modeling parameterization in climate models \cite{krasnopolsky2005new}. All these early works have utilized a simple feed-forward neural network with a single hidden layer. In the past decade, `deep learning' has made rapid progress in diverse areas such as speech translation \cite{sutskever2014sequence}, classification of images \cite{krizhevsky2017imagenet}, and playing the game of Go \cite{mnih2015human}. These applications can require a neural network with a large number of layers, special types of convolutional layers, and specialized architectures, such as generative adversarial networks \cite{goodfellow2014generative} along with the vast amount of data for the training. The machine learning (ML) methods can also be used to construct data-driven models of geophysical flows from the extensive data collected from satellites, in-situ scientific measurements, and futuristic internet-of-things (IoT) devices. Indeed, many researchers have explored the use of ML methods for earth science \cite{schneider2017earth,reichstein2019deep,mcgovern2017using,sonnewald2021bridging}.

One of the main challenges with ML includes the difficulty of incorporating existing domain knowledge and handling of uncertainty \cite{von2019informed,reichstein2019deep,kashinath2021physics}. The geophysical community utilizes the data assimilation (DA) framework to make use of physical laws and has robust ways of handling uncertainty in all parts of the problem \cite{carrassi2018data}. The DA framework is also capable of handling sparse and irregularly distributed observations gathered through satellite measurements \cite{eyre2020assimilation}. On the other hand, DA has mainly been employed for state estimation, such as the initial state of the system for the weather forecast. The forecast model can generally assumed to be perfect. However, the weather and climate models work on grid size around 10 km upwards, but the processes such as turbulence, radiation, and precipitation can take place at a scale around 1km or less. Hence, the parameterization scheme for these unresolved processes should take into account the average impact at the model grid scale. The ML can be applied to discover accurate parameterization schemes using the data from high resolution models \cite{rasp2018deep,gentine2018could,gagne2020machine}. The ML has also been utilized to replace the complete physical model of chaotic systems \cite{pathak2018model,vlachas2018data}. These data-driven models are computationally cheap and can be integrated within the DA framework for state estimation and forecasting \cite{tang2020deep}. 

The DA and ML share a lot of similarities as both approaches use the data to learn about the system using inverse methods \cite{hsieh1998applying,abarbanel2018machine,san2021hybrid} and the synergistic integration of the two is essential for learning improved models of the earth system \cite{bocquet2020bayesian,bonavita2020machine,farchi2020using,farchi2021comparison,geer2021learning}. One way to integrate the DA and ML is to build a hybrid neural-dynamical model via variational DA \cite{tang2001coupling}. This approach involves replacing the missing dynamics of the dynamical system with the neural network, and the parameters of the neural network were determined by the 4DVAR assimilation approach. The variational DA methods are commonly used to estimate the initial condition or model parameters, and are also implemented on operational weather and climate models \cite{navon1998practical,Zhu98impactof,courtier1998ecmwf,barker2004three}. Particularly, the backpropagation algorithm used to train the neural network has a strong mathematical \textcolor{rev1}{equivalence} with the adjoint method for calculating gradients in variational DA \cite{hsieh1998applying}, and this parallelism was utilized to determine parameters of the neural network to build a hybrid neural-dynamical model \cite{tang2001coupling}. Recently, the sequential data assimilation technique was applied to estimate the full state of the system using the truncated model (i.e., without accounting for unresolved processes), and then ML was utilized to learn the effect of unresolved part \cite{brajard2020combining}. \textcolor{rev2}{The work by Brajard et al. \cite{brajard2020combining} goes beyond high-resolution simulations and they develop ML-based parameterization using direct data in the realistic scenario of sparse and noisy observations.} The combination of ML-based parameterization to the physical core produces a hybrid model \cite{brajard2019representing}. The ML can be applied to learn the closure term accounting for the effect of subgrid-scale processes and the hybrid model can be corrected using sequential data assimilation in the online deployment phase \cite{pawar2020long,pawar2021data,ahmed2021closures}. Also, DA and ML can be applied iteratively to build a fully data-driven model from sparse and noisy observations \cite{BRAJARD2020101171}. The fully data-driven model can then be used for state estimation and forecasting. 

In this work, a hybrid modeling approach is presented with a data-driven representation for unknown physical processes and sequential data assimilation to correct the modeling error to achieve accurate analysis for chaotic dynamical systems. 
The paper is structured as follows. In Section~\ref{sec:hybrid}, the baseline Lorenz model, the Lorenz model with incomplete dynamics, \textcolor{rev1}{and the two-scale Lorenz model with data-driven parameterization} is presented along with the working of long short-term memory (LSTM) neural network. Section~\ref{sec:da} details the formulation of sequential data assimilation problem and lists the deterministic ensemble Kalman filter (DEnKF) algorithm. Section~\ref{sec:results} discusses the results of numerical experiments. \textcolor{rev3}{In particular, to test the proposed approach, we consider the following joint experiments: (i) incomplete physics-based model with LSTM for missing physics, (ii) incomplete physics-based model with LSTM for missing physics and DEnKF for data assimilation, and (iii) incomplete physics-based model and DEnKF for data assimilation.} Finally, the summary of this work and concluding remarks are provided in Section~\ref{sec:conclusion}. \textcolor{rev3}{Python codes for this work are available on GitHub \cite{GitHub}.}

\section{Hybrid neural-physics modeling} \label{sec:hybrid}

\subsection{Lorenz model} \label{sec:baseline}
\begin{figure}[h]
\centerline{\includegraphics[width=0.98\linewidth]{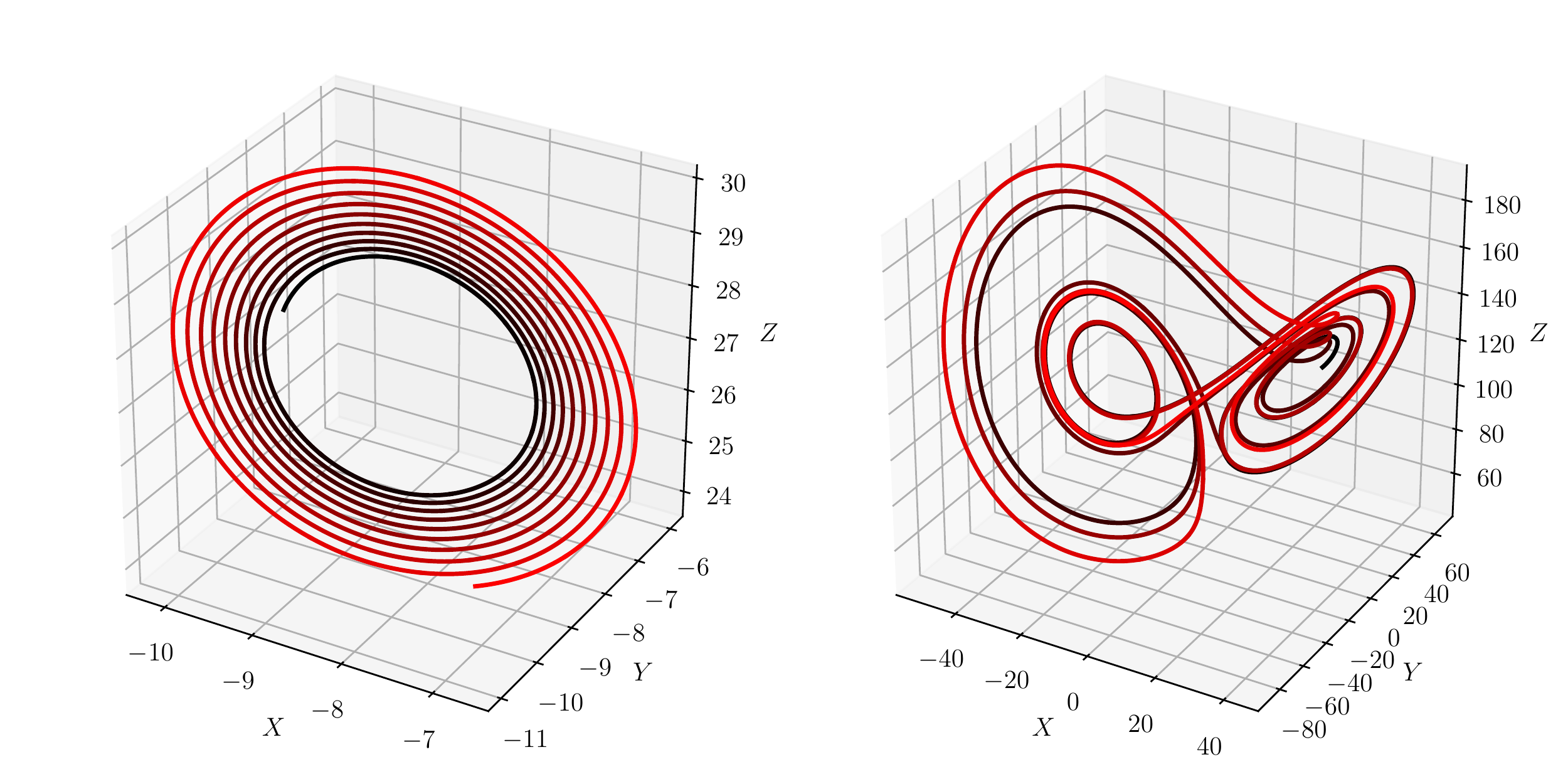}}
 \caption{The attractor of the Lorenz system integrated from time $t=0$ to $6$ s with the time step $\Delta t = 1 \times 10^{-3}$ s for the weakly nonlinear case (left) and the highly nonlinear case (right).}\label{fig:l63}
\end{figure}

We consider the following Lorenz system which is used as a simplified model for atmospheric convection \cite{lorenz1963deterministic}. The nonlinear system of three ordinary differential equations is 
\begin{align}
    \frac{dX}{dt} &= \sigma (Y-X), \label{eq:l63_x}\\
    \frac{dY}{dt} &= X (\rho - Z) - Y, \label{eq:l63_y}\\
    \frac{dZ}{dt} &= XY - \beta Z, \label{eq:l63_z} 
\end{align}
where $X,Y,Z$ are the state of the Lorenz system, and $\sigma,\rho,\beta$ are the system's parameter. In the above equations $X$ is proportional to the intensity of convection motion, $Y$ is proportional to the temperature difference in the horizontal direction, and $Z$ is proportional to the distortion of the temperature profile in the vertical direction. In the vector notation, the state of the Lorenz system can be defined as $\mathbf{x} = [X,Y,Z]$ and the Equations~\ref{eq:l63_x}-\ref{eq:l63_z} can be written as 
\begin{equation}
    \dot{\mathbf{x}} = \mathbf{f}(t,\mathbf{x}).
\end{equation}
The Lorenz system is integrated using the third-order Adams-Bashforth method \cite{durran1991third} as follows 
\begin{equation}\label{eq:ab3}
    \mathbf{x}^{(k+1)} = \mathbf{x}^{(k)} + \frac{\Delta t}{12} \large[ 23 \mathbf{f}^{(k-1)} - 16 \mathbf{f}^{(k-2)} +5 \mathbf{f}^{(k-3)}\large],
\end{equation}
\textcolor{rev1}{where $\mathbf{f}=[\sigma (Y-X),~X (\rho - Z) - Y,~ XY - \beta Z]$, and $\Delta t$ is the time step size}. The Lorenz system is very sensitive to the initial condition as well as the choice of parameters. Based on the selection of the parameters, the Lorenz system can exhibit either chaotic behavior or the transient chaotic behavior \cite{kaplan1979preturbulence}. Following the previous study \cite{tang2001coupling}, we investigate two cases here to evaluate the performance of the neural network in learning the chaotic dynamics. The first case is called the weakly nonlinear case, with the parameters $\sigma,~\rho,$ and $\beta$ set to 10, 28, and 8/3, respectively. The initial condition for the weekly nonlinear case is [-9.42, -9.43, 28.3] for $[X^{(0)},~Y^{(0)},~Z^{(0)}]$ \cite{gauthier1992chaos}. The Lorenz system displays near-regular oscillations with a gradually increasing amplitude in the devised integration period for the weakly nonlinear case. The initial condition for the highly nonlinear case is set to be $[X^{(0)},~Y^{(0)},~Z^{(0)}]$ equal to [22.8, 35.7, 114.9], and the parameters $\sigma,~\rho,$ and $\beta$ are set to 16.0, 120.1, and 4.0, respectively \cite{elsner1992nonlinear}. The dataset for training the neural network is generated with these reference solutions and is regarded as the `true' solution of the Lorenz equations. The Lorenz attractor for the weakly and highly nonlinear case is shown in Figure~\ref{fig:l63}. 

Next, we assume that the complete dynamics of the Lorenz model is unknown. Specifically, we assume that the evolution equation for the distortion of the temperature profile in the vertical direction, i.e., $Z$ is not available and must be approximated. The hybrid neural-physics Lorenz model composed of the physics-based equations for the known dynamics and the neural network to approximate the unknown dynamics can be written as 
\begin{align}
    \frac{dX}{dt} &= \sigma (Y-X), \label{eq:l63_x_np}\\
    \frac{dY}{dt} &= X (\rho - Z) - Y, \label{eq:l63_y_np}\\
    \frac{dZ}{dt} &= \mathcal{N}(\mathbf{x}(t),\cdots,\mathbf{x}(t-l\Delta t)), \label{eq:l63_z_np} 
\end{align}
where $l$ is the number of previous time steps used for the forecast of the variable $Z$ at next time step, i.e., $t+\Delta t$. 

\textcolor{rev1}{In this study, we train the neural network using the teacher forcing method \cite{williams1989learning}, which consists of using the ground truth data as the label for each training example. We assume that the data for the unknown state of the system (i.e., $Z$ variable in the Lorenz system) is either coming from measurements or from high-fidelity numerical simulations. The state derivative for the $Z$ variable at the $k$th time step is computed using the forward-difference scheme as follows}
\begin{equation}
    \frac{dZ}{dt} = \frac{Z^{(k+1)} - Z^{(k)}}{\Delta t}.
\end{equation}
Any type of machine learning model can be utilized to learn the relationship between the known and unknown dynamics. Tang et al. \cite{tang2001coupling} applied the feedforward neural network with a single hidden layer and five neurons to learn the missing physics in the Lorenz model. They observed that the neural network can approximate the unknown dynamics reasonably well for the weakly nonlinear case, but it fails for the highly nonlinear case. They proposed a better approach that makes use of the variational data assimilation where the dynamical constraints can be imposed for learning the parameters (i.e., weights and biases) of the neural network. Their proposed hybrid neural–dynamical variational data assimilation procedure leads to a very good prediction with almost the same forecast skill as the original Lorenz model. However, for the highly nonlinear case, the proposed method proposed a reasonable forecast only for few numerical experiments. This might have been due to the difficulty with the deterministic optimization algorithms within the variational assimilation in finding the global minima \cite{miller1994advanced, evensen1997advanced}. Even if the global optimum is found, the data in the assimilation window might be from one wing of the butterfly-shaped attractor, whereas the data in the forecast period lies on the second wing, thereby resulting in the poor forecast. This can be mitigated by applying stochastic data assimilation approaches like ensemble Kalman filter \cite{evensen1997advanced,pham2001stochastic,sakov2012iterative}. In this work, we revisit the problem of recovering missing physics in a dynamical system using the same examples as investigated by Tang et al. \cite{tang2001coupling} with the application of recurrent neural network (RNN) integrated within the deterministic ensemble Kalman filter algorithm \cite{sakov2008deterministic}.

\subsection{\textcolor{rev1}{Two-scale Lorenz model}}\label{sec:l2s}

\textcolor{rev1}{
The two-scale Lorenz model is given by a following set of ordinary differential equations (ODEs):
\begin{align}
    \frac{d X_n}{dt} &= -X_{n-1} (X_{n-2} - X_{n+1}) - X_n + F - \frac{hc}{b} \sum_{m=M(n-1) + 1}^{nM} Y_m, \label{eq:slow_true} \\
    \frac{d Y_m}{dt} &= -cbY_{m+1} (Y_{m+2} - Y_{m-1}) - c Y_m +   \frac{hc}{b} X_{\lfloor (m-1)/M \rfloor + 1} \label{eq:fast_true},
\end{align}
where $n = 1,\dots, N$, $ m = 1,\dots, MN$, and $\lfloor \cdot \rfloor$ is the modulo operation. \textcolor{rev2}{The $X$ and $Y$ variables are periodic, i.e., $X_1=X_N$ and $Y_1=Y_{MN}$.} In the present study, the number of $X$ variables is set at $N=8$ and the number of $Y$ variables per $X$ variables is set to $M=32$. We also set the coupling constant between $X$ and $Y$ variables to $h=1$, the spatial-scale ratio to $b=10$, and the temporal-scale ratio to $c=10$. The forcing term is set at a large value of $F=20$ to ensure chaotic behavior. These parameters are such that one model time unit (MTU) is approximately equivalent to five atmospheric days \cite{arnold2013stochastic}. Recently, this model with the same parameter settings was applied to test the feasibility of generative adversarial networks to learn stochastic parameterization in multiscale systems \cite{gagne2020machine}.  
}

\textcolor{rev1}{
The two-scale Lorenz model given by Equation~\ref{eq:slow_true} and Equation~\ref{eq:fast_true} is regarded as the truth model that must be simulated. In weather and climate modeling, the governing equations of motion of the system are known. However, it is not possible to resolve every small scale due to limited computational resources. Therefore, the effect of small scale variables are typically parameterized as a function of resolved variables. A forecast model for the two-scale Lorenz equations with the parameterization of the small $Y$ scale variables on the resolved $X$ variables can be written as
\begin{equation}
    \frac{d \Tilde{X}_n}{dt} = -\Tilde{X}_{n-1} (\Tilde{X}_{n-2} - \Tilde{X}_{n+1}) - \Tilde{X}_n + F - G_n, \label{eq:slow}
\end{equation}
where $\Tilde{X}_n$ is the forecast estimate of ${X}_n$ and $G_i$ represent the effect of unresolved variables. The parameterization is usually a function of the resolved variables, i.e., $\Tilde{\mathbf{X}} \in \mathbb{R}^n$, and can be written mathematically as 
\begin{equation}
    \frac{hc}{b}\sum_{m=N(n-1) + 1}^{nM} Y_m :\approx ~ G_n =~\mathfrak{N}(\Tilde{\mathbf{X}}),
\end{equation}
where $\mathfrak{N}(\cdot)$ is the nonlinear mapping for representing the effect of unresolved scales on resolved scale variables. This mapping can be based on physical arguments or some prior information about subgrid scale processes. Alternatively, data-driven models can be used to exploit the data generated from experimental measurements or high-fidelity numerical simulations to learn the relation between resolved and unresolved scales. Indeed, in recent years, data-driven models have been successfully applied for parameterizing subgrid scale processes in complex geophysical flows \cite{rasp2018deep,gentine2018could,gagne2020machine,pawar2021data,maulik2019subgrid}. In the present study, we utilize the LSTM neural network to learn the subgrid scale parameterization of unresolved scales. Once we have the ground truth data available for large scale $X$ variables, we can compute the \emph{true} subgrid scale parameterization as follows
\begin{equation} \label{eq:two_scale_g}
    G_n^{(k)} = -X_{n-1}^{(k)} (X_{n-2}^{(k)} - X_{n+1}^{(k)}) - X_n^{(k)} + F - \bigg(\frac{X_n^{(k+1)} - X_n^{(k)}}{\Delta t} \bigg),
\end{equation}
where $\Delta t$ is the time step of integration of the forecast model. The ground truth data for training the neural network is generated by temporally integrating the two-scale Lorenz system using the fourth-order Runge-Kutta (RK4) time stepping scheme.
}

\subsection{Long short-term memory neural network} \label{sec:lstm}
In this work, we apply the long short-term memory (LSTM) neural network to predict the missing physics (i.e., incomplete dynamics) from the resolved physics of the dynamical system. The motivation behind applying LSTM is due to its success in modeling high-dimensional spatio-temporal chaotic time series of physical systems \cite{pathak2018model,vlachas2018data,vlachas2020backpropagation,maulik2020non}. LSTM is a type of recurrent neural network (RNN) that can capture the long-term dependencies in the evolution of time series data \cite{hochreiter1997long}. RNNs contain loops that allow them to persist information from one time step to another and can be expressed as
\begin{align}
    \mathbf{h}^{(t)} &= f_{h \rightarrow h} (\mathbf{o}^{(t)}, \mathbf{h}^{(t-1)}), \\
    \tilde{\mathbf{o}}^{(t+1)} &= f_{h \rightarrow o} (\mathbf{h}^{(t)}),
\end{align}
where $\mathbf{h}^{(t)} \in \mathbb{R}^{d_h}$ is the hidden state at time $t$, $\mathbf{o}^{(t)} \in \mathbb{R}^{d_h}$ is the input vector at time $t$, $f_{h \rightarrow h}$ is the hidden to hidden mapping, and $f_{h \rightarrow o}$ is the hidden to output mapping. The output of the model is the forecast $\tilde{\mathbf{o}}^{(t+1)}$ at time step $t+1$. 

\begin{figure}[h]
\centerline{\includegraphics[width=0.98\linewidth]{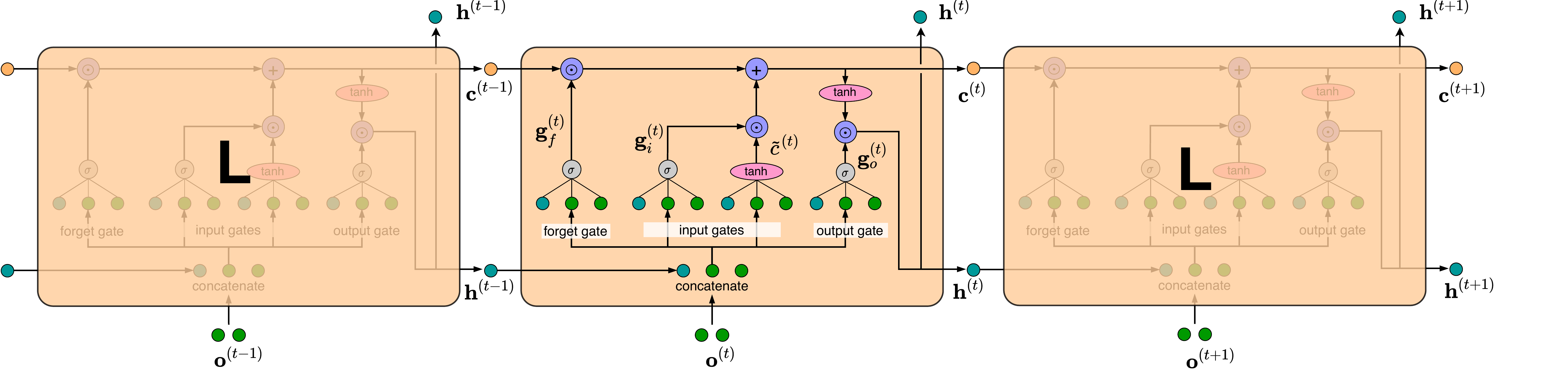}}
 \caption{Schematic of the unfolded long short-term memory (LSTM) neural network architecture. The LSTM uses the gating mechanism that allows forgetting and storing of information in the processing of the hidden state. The L is used to denote the LSTM cell.}\label{fig:lstm}
\end{figure}

One of the limitations of RNNs is vanishing (or exploding) gradient to capture the long-term dependencies. This stems from the fact that the gradient is multiplied with the weight matrix repetitively during backpropagation through time (BPTT). The LSTM mitigates the issue with vanishing (or exploding) gradient by employing the gating mechanism that allows information to be forgotten. The equations that implicitly define the mapping from hidden state from the previous time step (i.e., $\mathbf{h}^{(t-1)}$) and input vector at the current time step (i.e., $\mathbf{o}^{(t)}$) to the forecast hidden state (i.e., $\mathbf{h}^{(t)}$) can be written as
\begin{align}
    \mathbf{g}^{(t)}_f &= \sigma(\mathbf{W}_f [\mathbf{h}^{(t-1)},\mathbf{o}^{(t)}] + \mathbf{b}_f), \\
    \mathbf{g}^{(t)}_i &= \sigma(\mathbf{W}_i [\mathbf{h}^{(t-1)},\mathbf{o}^{(t)}] + \mathbf{b}_i), \\
    \tilde{\mathbf{c}}^{(t)} &= \text{tanh}(\mathbf{W}_c [\mathbf{h}^{(t-1)},\mathbf{o}^{(t)}] + \mathbf{b}_c), \\
    \mathbf{c}^{(t)} &= \mathbf{g}^{(t)}_f \odot \mathbf{c}^{(t-1)} + \mathbf{g}^{(t)}_i \odot \tilde{\mathbf{c}}^{(t)}, \\
    \mathbf{g}^{(t)}_o &= \sigma(\mathbf{W}_o [\mathbf{h}^{(t-1)},\mathbf{o}^{(t)}] + \mathbf{b}_o), \\
    \mathbf{h}^{(t)} &= \mathbf{g}^{(t)}_o \odot \text{tanh} (\mathbf{c}^{(t)}),
\end{align}
where $\mathbf{g}^{(t)}_f,~\mathbf{g}^{(t)}_i,~\mathbf{g}^{(t)}_o \in \mathbb{R}^{d_h}$ are the forget gate, input gate, and output gate, respectively. The $\mathbf{o}^{(t)} \in \mathbb{R}^{d_i}$ is the input vector at time $t$, $\mathbf{h}^{(t)} \in \mathbb{R}^{d_h}$ is the hidden state, $\mathbf{c}^{(t)} \in \mathbb{R}^{d_h}$ is the cell state, $\mathbf{W}_f,~\mathbf{W}_i,~\mathbf{W}_c,~\mathbf{W}_o \in \mathbb{R}^{d_h \times (d_h + d_i)}$ are the weight matrices, and $\mathbf{b}_f,~\mathbf{b}_i,~\mathbf{b}_c,~\mathbf{b}_o \in \mathbb{R}^{d_h}$ are the bias vectors. The symbol $\odot$ denotes the element-wise multiplication, and $\sigma$ is the sigmoid activation function. The weights and biases are optimized using the BPTT algorithm \cite{werbos1990backpropagation}. The above set of equations are unfolded to capture the temporal dependencies in predicting future state $\mathbf{o}^{(t+1)}$ given $\mathbf{o}^{(t)},\mathbf{o}^{(t-1)},\cdots,\mathbf{o}^{(t-l)}$. The $l$ is referred to as the lookback which governs how much amount of the old temporal information is needed to forecast the future state of the system accurately. An illustration of the gating mechanism in the LSTM cell is given in Figure~\ref{fig:lstm}. 

\textcolor{rev1}{For the Lorenz system, the LSTM is trained to predict the state derivative $\dot{Z}$ at the $k$th time step using the temporal history of the full state of the system for $l$ past consecutive states, i.e., $\{\mathbf{x}^{(k)},\mathbf{x}^{(k-1)},\dots,\mathbf{x}^{(k-l+1)}\}$. Since we are using the information of only $l$ past temporally consecutive states as the input, the LSTM can capture dependencies up to $l$ previous time steps.} \textcolor{rev2}{The inspiration for exploiting the recent history of state variables to predict the state derivative $dZ/dt$ comes from Taken's theorem \cite{takens1981detecting} and several other studies on the use of RNNs for chaotic dynamical systems \cite{cechin2008optimizing,dubois2020data,gers2002applying,vlachas2018data}.}
Once the LSTM is trained, it is used to iteratively predict the future state in an auto-regressive manner. In this method, the initial condition for the first $(l)$ time steps (equal to lookback) is provided. This information is used to predict the forecast state at $(l+1)$th time step. Then the state of the system from $(2)-(l+1)$ is used to predict the forecast state at $(l+2)$th time step. This procedure is continued until the final time step. \textcolor{rev1}{Since the inference stage is different from the training with teacher forcing method,} there will be an accumulation of modeling error in the auto-regressive deployment of the trained LSTM network and this causes the deviation in the predicted trajectory of the hybrid neural-physics model \cite{bengio2015scheduled}. \textcolor{rev1}{One of the remedies to avoid this discrepancy between training and inference procedure is to adopt the training without teacher forcing \cite{sangiorgio2020robustness}. In the training without teacher forcing method, the prediction of the LSTM network is fed back into the input features for future prediction. Training the LSTM without the teacher forcing method has been shown to improve the accuracy and robustness for chaotic attractor \cite{sangiorgio2020robustness}.} Another method of error correction is through exploiting the online measurements within the sequential data assimilation framework for attaining accurate forecast over a longer period for chaotic dynamical systems. 

\textcolor{rev1}{For the two-scale Lorenz model, the input of the LSTM network is the full state of the system for resolved $X$ variables for $l$ past temporally consecutive states and the output is the subgrid scale parameterization at the $k$th time step. Mathematically, the LSTM mapping for learning parameterizations in a two-scale Lorenz model can be represented as
\begin{equation}
    \{\mathbf{X}^{(k)}, \mathbf{X}^{(k-1)}, \dots, \mathbf{X}^{(k-l+1)}\} \in \mathbb{R}^{l \times N} \rightarrow \{\mathbf{G}^{(k)}\} \in \mathbb{R}^N.
\end{equation}
where $\mathbf{X}$ is the resolved flow variables and $\mathbf{G}$ is true parameterization.
}

\section{Sequential data assimilation} \label{sec:da}
Data assimilation (DA) is a technique to incorporate sparse and irregularly distributed noisy measurements with the physical model of the system to achieve an accurate prediction of the state trajectory. DA is generally classified into two categories: variational and sequential DA. In variational DA, all observations over a particular time window are utilized to minimize the cost function with certain constraints to compute the model trajectory that best fits the observations. The variational DA is particularly well suited for re-analyses problems to obtain the best possible state of the system at some time $t$ using observations before and after this time \cite{navon2009data}.  In sequential DA, the system's state is evolved in time using background information (i.e., the physical model) until observations become available. At this instant, an update (correction) to the system's state is determined and the solver is re-initialized with this analyzed state. This procedure is continued until new observations get available, and so on. There is a rich literature on DA \cite{lewis2006dynamic,simon2006optimal,evensen2009data} and here we discuss only sequential DA problem and then outline the algorithm procedure for the deterministic ensemble Kalman filter (DEnKF).

We consider the dynamical system whose evolution can be represented as 
\begin{equation}\label{eq:dyn_model}
    \mathbf{x}^{(k+1)} = \textbf{M}_{t_k \rightarrow t_{k+1}}(\mathbf{x}^{(k)}) + \mathbf{w}^{(k+1)},
\end{equation}
where $\mathbf{x}^{(k)} \in \mathbb{R}^n$ is the state of the system at discrete time $t_k$, and $\textbf{M}:\mathbb{R}^n \rightarrow \mathbb{R}^n$ is the nonlinear model operator that defines the evolution of the system. The term $\mathbf{w}^{(k+1)}$ denotes the model error that takes into account any type of uncertainty in the model that can be attributed to boundary conditions, imperfect models, etc. Let $\mathbf{z}^{(k)} \in \mathbb{R}^m$ be observations of the state vector obtained from sparse, noisy measurements and can be written as
\begin{equation}
    \mathbf{z}^{(k+1)} = \mathbf{q}(\mathbf{x}^{(k+1)}) + \mathbf{v}^{(k+1)},
\end{equation}
where $\mathbf{q}(\cdot)$ is a nonlinear function that maps $\mathbb{R}^n \rightarrow \mathbb{R}^m$, and $\mathbf{v}^{(k+1)} \in \mathbb{R}^m$ is the measurement noise. We assume that the measurement noise is a white Gaussian noise with zero mean and the covariance matrix $\mathbf{R}^{(k+1)}$, i.e., $\mathbf{v}^{(k+1)} \sim {\cal{N}}(0,\mathbf{R}^{(k+1)})$. Additionally, the noise vectors $\mathbf{w}^{(k+1)}$ and $\mathbf{v}^{(k+1)}$ are assumed to be uncorrelated to each other at all time steps. The sequential DA can be considered as a problem of estimating the state $\mathbf{x}^{(k+1)}$ of the system given the observations up to time $t_{k+1}$, i.e., $\mathbf{z}^{(1)},\dots,\mathbf{z}^{(k+1)}$. When  we utilize observations to estimate the state of the system, we say that the data are assimilated into the model, and use the notation $\widehat{\mathbf{x}}^{(k+1)}$ to denote an analyzed state estimate of the system at time $t_{k+1}$. When all the observations before (but not including) time $t_{k+1}$ are applied for estimating the state of the system, then we call it the forecast estimate and denote it as $\mathbf{x}^{(k+1)}_f$.

The ensemble Kalman filter (EnKF) \cite{burgers1998analysis} follows the Monte Carlo estimation method to approximate the covariance matrix in the Kalman filter equations \cite{kalman1960new}. Inflation and covariance localization approaches have been often used in the EnKF framework to mitigate small number of ensembles \cite{houtekamer2005ensemble,raanes2019adaptive,attia2019dates,ahmed2020pyda}. Instead of modeling the exact evolution of a probability density function under nonlinear dynamics, ensemble methods maintain an empirical approximation to the target distribution in the form of a set of ensemble members $\mathbf{\widehat{X}}^{(k)}(i)$ for $i=1 \dots N$. We begin by initializing the state of the system for different ensemble members $\mathbf{\widehat{X}}^{(0)}(i)$ drawn from the distribution ${\cal{N}} (\mathbf{\widehat{x}}^{(0)},\mathbf{P}^{(0)})$, where $\mathbf{\widehat{x}}^{(0)}$ represents the best-known state estimate at time $t_0$, and $\mathbf{P}^{(0)}$ is the initial covariance error matrix.

The propagation of the state for each ensemble member over the time interval $[t_k,t_{k+1}]$ can be written as
\begin{align}
\mathbf{X}^{(k+1)}_f(i) &= \mathbf{M}_{t_k \rightarrow t_{k+1}}(\widehat{\mathbf{X}}^{(k)}(i)) + \mathbf{w}^{(k+1)}.
\label{eq:enkf_xk}
\end{align}
The term $\mathbf{w}^{(k+1)}$ accounting for model imperfections is usually assumed to be Gaussian noise. The prior state and the covariance matrix are approximated using the sample mean and error covariance matrix $\mathbf{P}^{(k+1)}_f$ as follows
\begin{align}
    \mathbf{x}^{(k+1)}_f &= \frac{1}{N} \sum_{i=1}^N \mathbf{X}^{(k+1)}_f(i), \label{eq:enkf_xf}\\
\mathbf{A}^{(k+1)}_f(i) &= \mathbf{X}^{(k+1)}_f(i) - \mathbf{x}^{(k+1)}_f, \\
\mathbf{P}^{(k+1)}_f &= \frac{1}{N-1} \sum_{i=1}^N  \mathbf{A}^{(k+1)}_f(i) (\mathbf{A}^{(k+1)}_f(i))^{\text{T}}, \label{eq:enkf_pf}
\end{align}
where the superscript $\text{T}$ denotes the transpose, \textcolor{rev1}{and $\mathbf{A}^{(k+1)}_f(i)$ is the anomalies between the forecast estimate for the $i$th ensemble and the sample mean.} Once the observations gets available at time $t_{k+1}$, the forecast state estimate is assimilated using the Kalman filter analysis equation as follows 
\begin{align}
    \widehat{\mathbf{x}}^{(k+1)} = \mathbf{x}^{(k+1)}_f + \mathbf{K}^{(k+1)}[\mathbf{z}^{(k+1)} - \mathbf{q}(\mathbf{x}^{(k+1)}_f)]. \label{eq:denkf_xa}
\end{align}
In contrast to the ensemble Kalman filter algorithm, the DEnKF does not employ any perturbed observations. The Kalman gain matrix is computed using its square root version (without storing or computing $\mathbf{P}^{(k+1)}_f$ explicitly) as follows
\begin{multline}
    \mathbf{K}^{(k+1)} = \frac{\mathcal{A}^{(k+1)}_f(\mathbf{Q}^{(k+1)} \mathcal{A}^{(k+1)}_f)^{\text{T}}}{N-1} \bigg[\frac{(\mathbf{Q}^{(k+1)}\mathcal{A}^{(k+1)}_f)(\mathbf{Q}^{(k+1)}\mathcal{A}^{(k+1)}_f)^\text{T}}{N-1}\\
    + \mathbf{R}^{(k+1)} \bigg]^{-1},
\end{multline}
where $\mathbf{Q} \in \mathbb{R}^{m \times n}$ is the Jacobian of the observation operator (i.e., $Q_{kl} = \frac{\partial q_k}{ \partial x_l}$), and the matrix $\mathcal{A}^{(k+1)}_f \in \mathbb{R}^{n \times N}$ is concatenated as follows
\begin{align}
\mathcal{A}^{(k+1)}_f = [\mathbf{A}^{(k+1)}_f(1), \mathbf{A}^{(k+1)}_f(2), \dots, \mathbf{A}^{(k+1)}_f(N)].
\end{align}
The anomalies for all ensemble members are then updated separately with half the Kalman gain as shown below 
\begin{align}
    \widehat{\mathbf{A}}^{(k+1)}(i) = \mathbf{A}^{(k+1)}_f(i) - \frac{1}{2}\mathbf{K}^{(k+1)} \mathbf{Q}^{(k+1)}\mathbf{A}^{(k+1)}_f(i). \label{eq:denkf_aa}
\end{align}
The state for all ensemble members is updated by adding ensemble anomalies to analysis state estimate and can be written as 
\begin{align}
    \widehat{\mathbf{X}}^{(k+1)}(i) = \widehat{\mathbf{x}}^{(k+1)} + \lambda \cdot \widehat{\mathbf{A}}^{(k+1)}(i), \label{eq:denkf_ea}
\end{align}
where $\lambda$ is the inflation factor to account for modeling errors. The above ensembles are used as initial ensembles for the next assimilation cycle and the procedure is continued. 

\section{Results and discussions} \label{sec:results}
\begin{figure}[h]
\centerline{\includegraphics[width=0.98\linewidth]{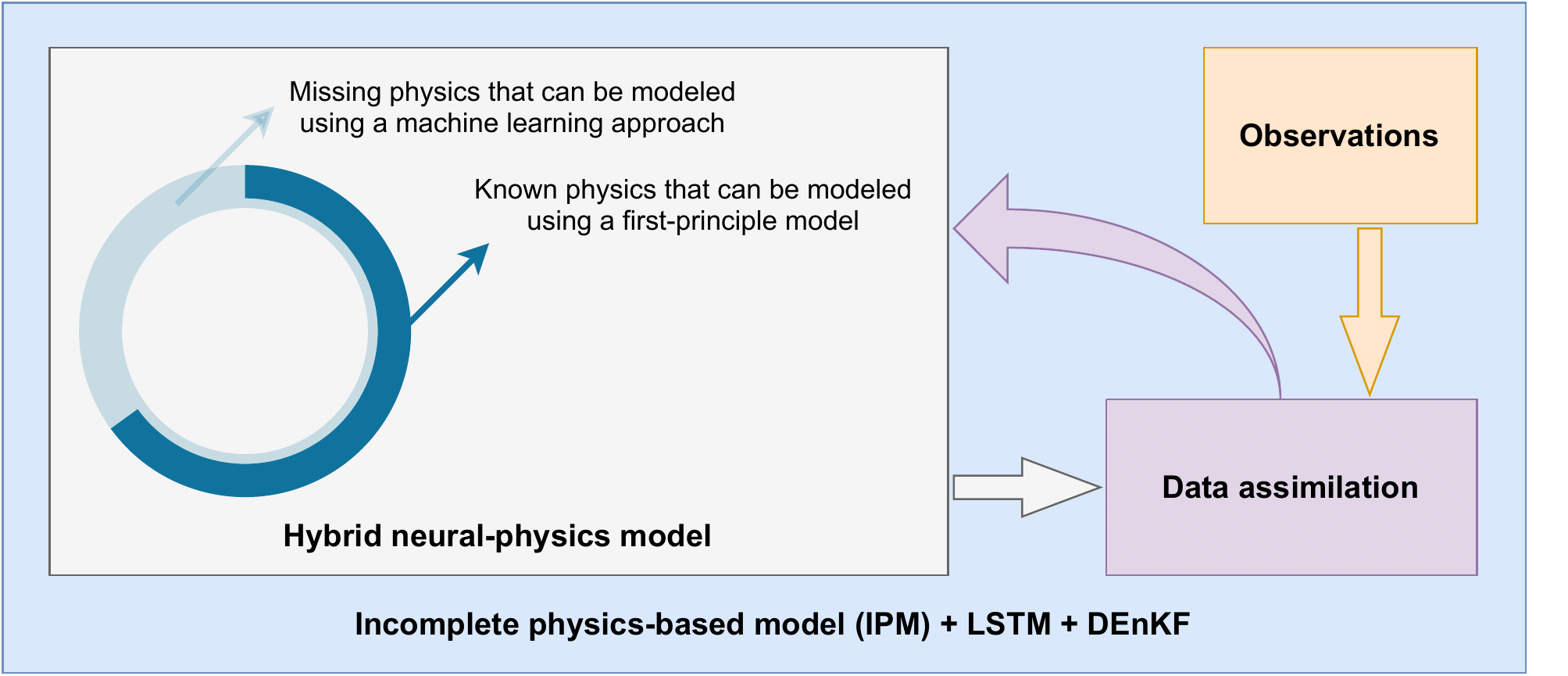}}
 \caption{\textcolor{rev3}{The general structure of the proposed hybrid neural-physics modeling approach. Machine learning is used to learn the missing dynamics or parameterization of unresolved processes and this provides us with the hybrid neural-physics model. The hybrid neural-physical model can be further corrected through the data assimilation step making use of sparse and noisy observations in the online deployment phase.}}\label{fig:def}
\end{figure}

\textcolor{rev3}{The general structure of the proposed hybrid neural-physics modeling approach is illustrated in Figure~\ref{fig:def}. We stress here that for many physical systems the complete description for dynamics of the system is not available and the missing physics for such systems can be modeled using machine learning tools. This provides us with the hybrid neural-physics model. The forecast of the hybrid neural-physics model can be further corrected using the sparse and noisy observations through data assimilation (specifically, we utilize the DEnKF algorithm). We consider a series of numerical experiments: (i) incomplete physics-based model with LSTM for missing physics (without data assimilation), (ii) incomplete physics-based model with LSTM for missing physics and DEnKF for data assimilation, and (iii) incomplete physics-based model (without LSTM closure) and DEnKF for data assimilation. We emphasize that we apply machine learning to improve the model of the system (e.g., see Figure~\ref{fig:def}). Data assimilation is then used in the analysis step to improve the forecast of the system.}


\subsection{\textcolor{rev1}{Lorenz model}}
\subsubsection{Hybrid neural-physics model} \label{sec:hybrid_l63}
\begin{figure}[h]
\centerline{\includegraphics[width=0.98\linewidth]{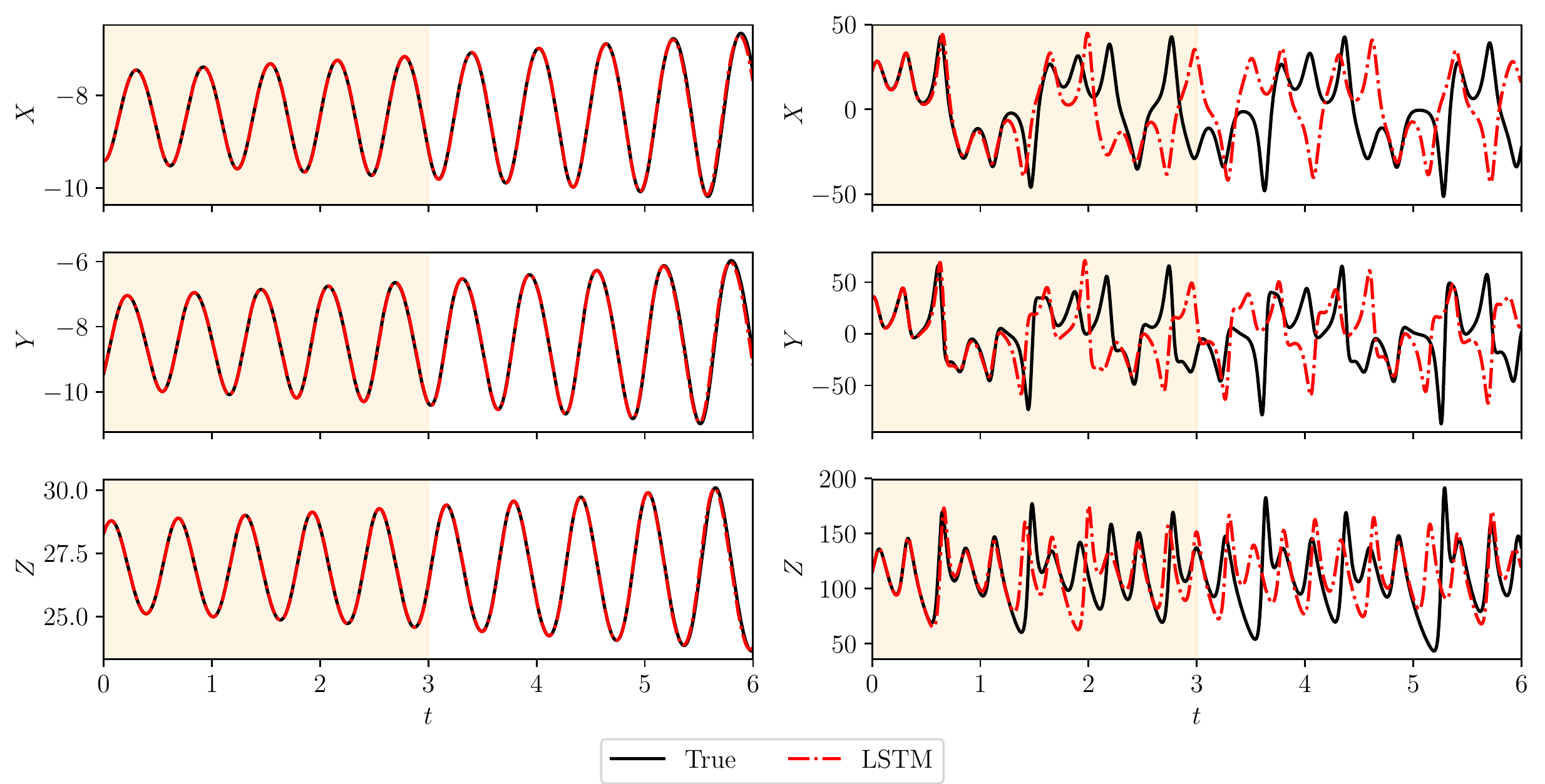}}
 \caption{The trajectory of the Lorenz system computed using the true physics-based model and the hybrid model where the $dZ/dt$ is approximated using the neural network for the weakly nonlinear case (left) and the highly nonlinear case (right). \textcolor{rev3}{Here, LSTM refers to forecast with the hybrid model.}}\label{fig:l63hybrid}
\end{figure}

The Lorenz model is integrated using the third-order Adams-Bashforth method with the time step $\Delta t = 1 \times 10^{-3}$ for 6000 time steps, i.e., $t=6$. The LSTM network is trained using the data from time $t=0$ to $t=3$. For training the LSTM network, we apply the lookback $l=6$ meaning that the state of the system for six previous time steps is used to predict the unknown dynamics (i.e., $dZ/dt$ in our case) at the next time step. We use the full state of the system $[X,Y,Z]$ as the input features to the LSTM network. During the deployment of the trained LSTM network, the variable $Z$ at the next time step is calculated using Equation~\ref{eq:ab3} with the predicted $dZ/dt$. Since the LSTM network is deployed in an auto-regressive method, there is an accumulation of errors from one time step to another. The chaotic systems are very sensitive to the initial condition and this makes the accurate forecast of chaotic systems difficult over a longer period. Figure~\ref{fig:l63hybrid} displays the prediction of the Lorenz system using the true model and the hybrid neural-physics model for the weakly and highly nonlinear case. The hybrid neural-physics model can predict the correct dynamics of the $Z$ variable and the predicted state trajectory is very close to the true state of the system. However, for the highly nonlinear Lorenz system, the predicted state trajectory starts deviating from the true state trajectory at around $t \approx 1.5$. \textcolor{rev1}{The largest Lyapunov exponent of the highly nonlinear Lorenz system is 2.33. Therefore, the time $t \approx 1.5$ corresponds to 3.5 Lyapunov time ($\Lambda_1^{-1}$, where $\Lambda_1$ is the largest Lyapunov exponent of the Lorenz system).} Even though the hybrid neural-physics model can predict the switching between lobes, the predicted state trajectory is considerably different from the true state trajectory for the highly nonlinear case. Therefore, the online data should be integrated using sequential DA to achieve an accurate forecast by limiting the accumulation of prediction errors.  

\subsubsection{Hybrid neural-physics model coupled with DA}
\begin{figure}[h]
\centerline{\includegraphics[width=0.98\linewidth]{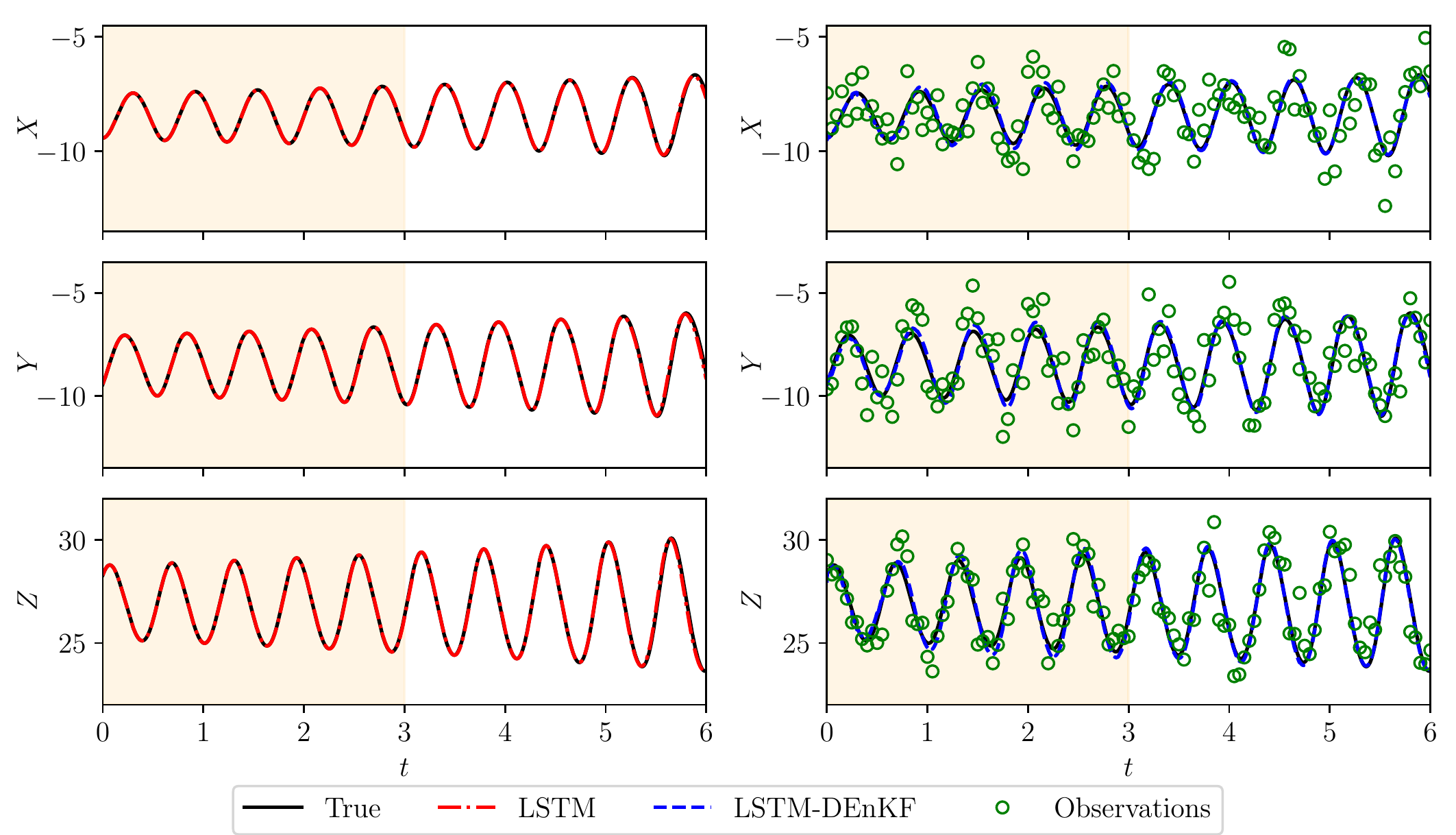}}
 \caption{The trajectory of the weakly nonlinear Lorenz system computed using the true physics-based model, the hybrid model where the $dZ/dt$ is approximated using the neural network, and the assimilated results with the hybrid model. The observation noise is drawn from $\mathcal{N}(0,1)$, and are collected after every 50 time steps, i.e., the time interval between two observations is 0.05. \textcolor{rev3}{Here, LSTM refers to forecast with the hybrid model and LSTM-DEnKF refers to an analysis state of the system.}}\label{fig:l63whybrid50}
\end{figure}

\begin{figure}[h]
\centerline{\includegraphics[width=0.98\linewidth]{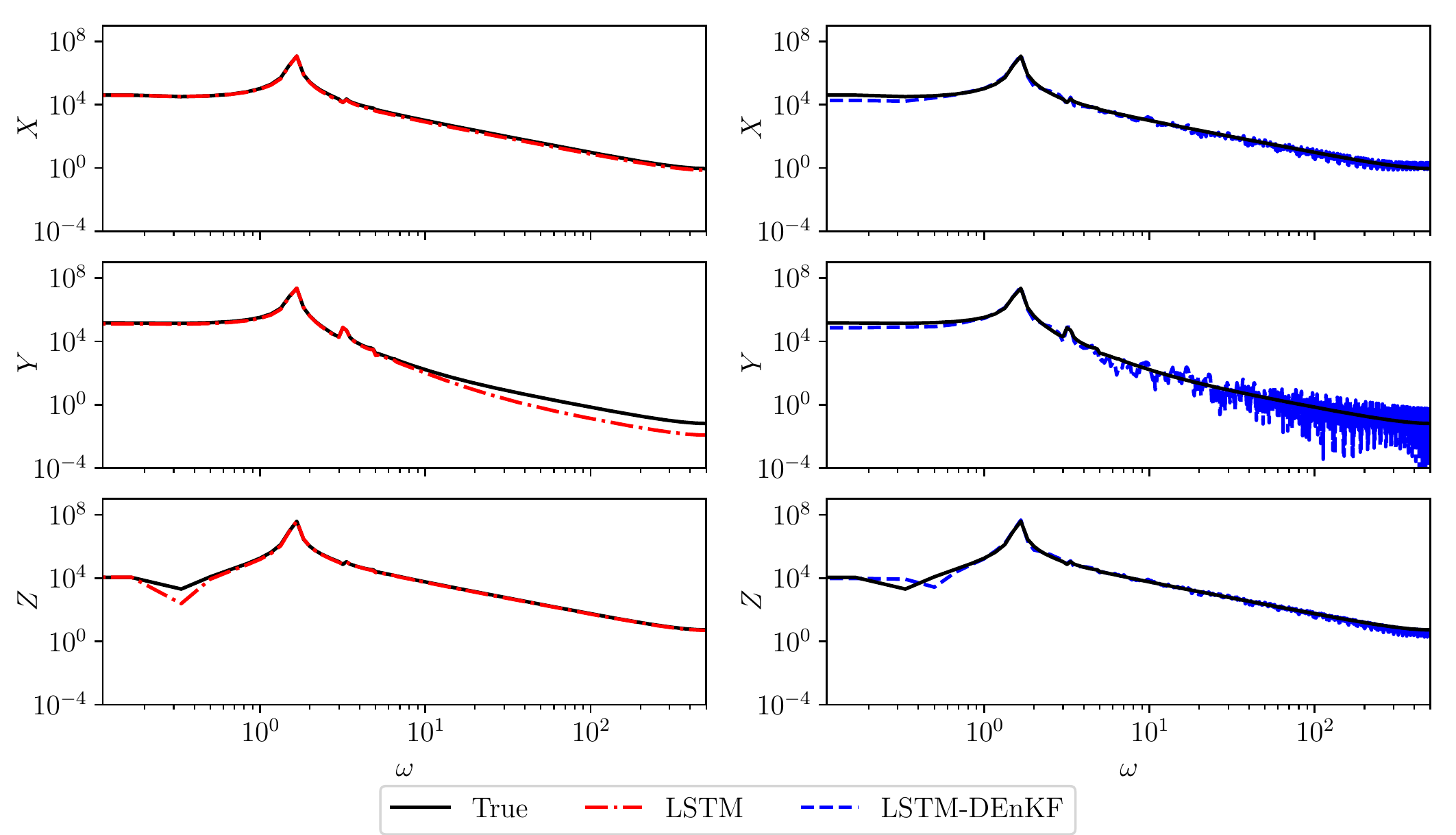}}
 \caption{Power spectra analysis of the weakly nonlinear Lorenz system computed using the true physics-based model, the hybrid model where the $dZ/dt$ is approximated using the neural network, and the assimilated results with the hybrid model. The observation noise is drawn from $\mathcal{N}(0,1)$, and are collected after every 50 time steps, i.e., the time interval between two observations is 0.05. \textcolor{rev3}{Here, LSTM refers to forecast with the hybrid model and LSTM-DEnKF refers to an analysis state of the system.}}\label{fig:l63psdw}
\end{figure}

\begin{figure}[h]
\centerline{\includegraphics[width=0.98\linewidth]{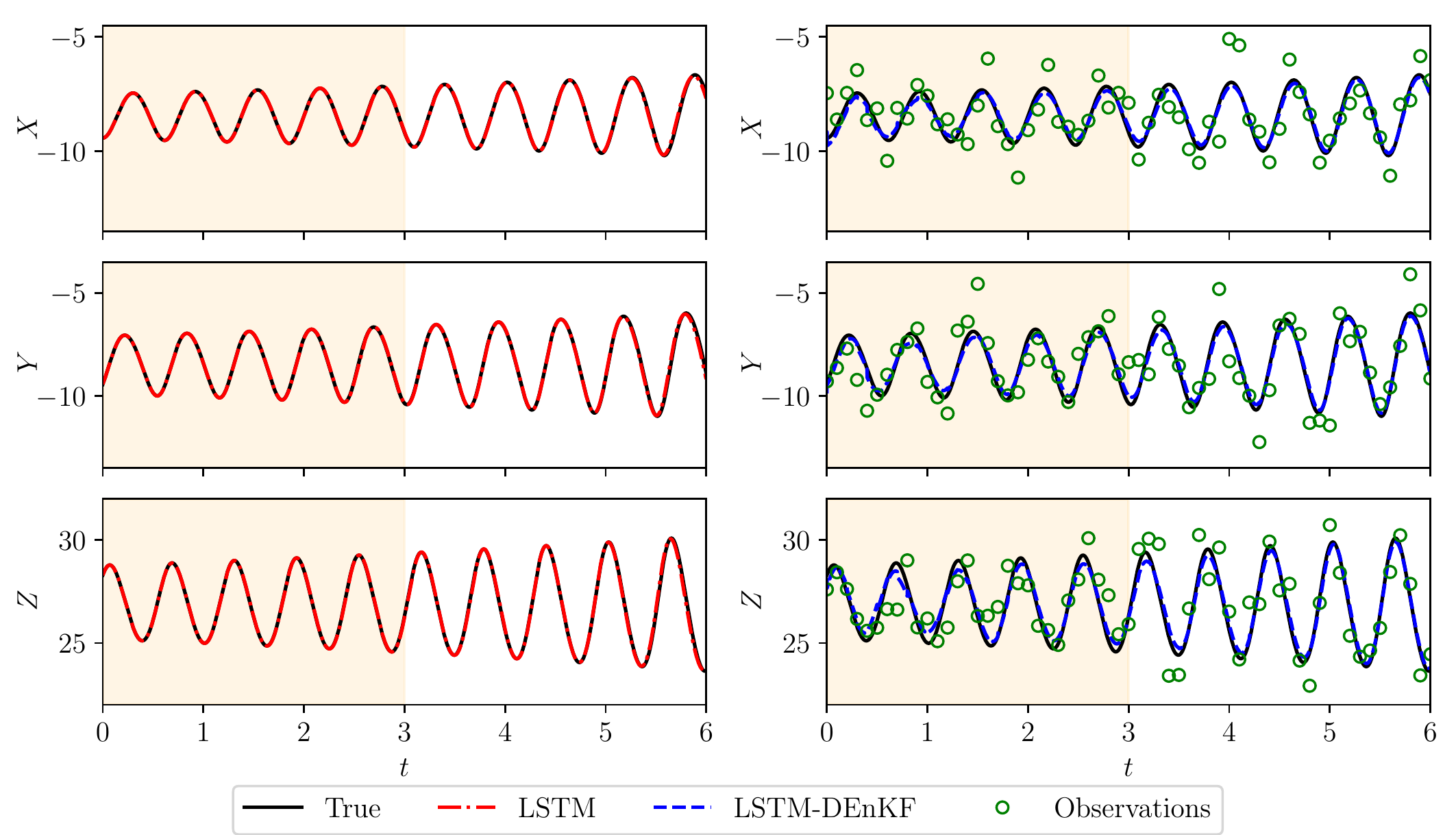}}
 \caption{The trajectory of the weakly nonlinear Lorenz system computed using the true physics-based model, the hybrid model where the $dZ/dt$ is approximated using the neural network, and the assimilated results with the hybrid model. The observation noise is drawn from $\mathcal{N}(0,1)$, and are collected after every 100 time steps, i.e., the time interval between two observations is 0.1. \textcolor{rev3}{Here, LSTM refers to forecast with the hybrid model and LSTM-DEnKF refers to an analysis state of the system.}}\label{fig:l63whybrid100}
\end{figure}

\begin{figure}[h]
\centerline{\includegraphics[width=0.98\linewidth]{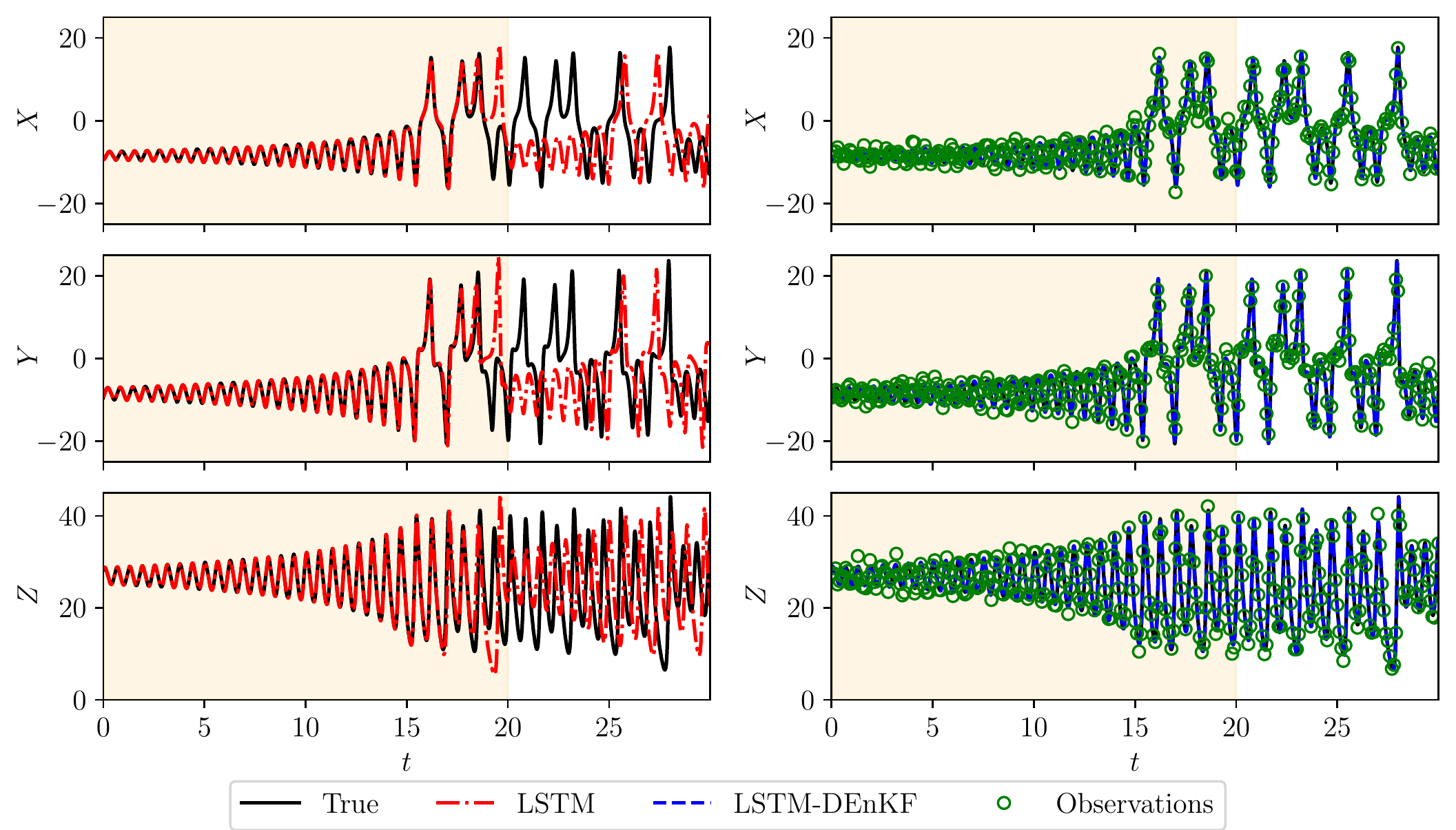}}
 \caption{\textcolor{rev3}{The trajectory of the weakly nonlinear Lorenz system  for a long time computed using the true physics-based model, the hybrid model where the $dZ/dt$ is approximated using the neural network, and the assimilated results with the hybrid model. The observation noise is drawn from $\mathcal{N}(0,1)$, and are collected after every 100 time steps, i.e., the time interval between two observations is 0.1. Here, LSTM refers to forecast with the hybrid model and LSTM-DEnKF refers to an analysis state of the system.}}\label{fig:l63whybrid100_long}
\end{figure}

Next, we show the performance of the DEnKF algorithm that uses the hybrid neural-physics model as the forward model and compare it with only using the hybrid neural-physics model. Figure~\ref{fig:l63whybrid50} shows the trajectory of the weakly nonlinear Lorenz system where the observations are assimilated after every 50 time steps. The observations are obtained by adding the Gaussian noise with the variance $\sigma_b^2=1.0$ and we assume that the full state is observable. We initialize 10 different ensemble members by adding the Gaussian noise with the variance $\sigma_0^2=1.0$ to the true state of the system at time $t=0$. From Figure~\ref{fig:l63whybrid50}, we can notice that the prediction with the hybrid neural-physics model integrated within the DEnKF algorithm matches almost exactly with the true trajectory of the weakly nonlinear Lorenz system. This is to be expected as even the hybrid neural-physics model was able to predict the state trajectory close to the true trajectory of the weakly nonlinear Lorenz system and ensemble Kalman filter approaches are robust for sparse and noisy observations in improving the state estimation. In Figure~\ref{fig:l63psdw}, we compare the power spectral density for all three states of the true model, hybrid neural-physics model, and the hybrid neural-physics model coupled with DEnKF. The power spectral density (PSD) is calculated as the square of the absolute value of the discrete Fourier transform of the time series of interest and can be written as 
\begin{equation}
    \text{PSD} = |\widehat{{u}}(\omega)|^2,
\end{equation}
where $\widehat{{u}}(\omega)$ is the Fourier transform of a time series. We observe some noise in the PSD for the $Y$ variable, especially at higher frequencies which can be attributed to the Gaussian noise of observations that are assimilated for the state estimation in the DEnKF algorithm. We further increase the time interval at which the observations are assimilated to 100 time steps. As shown in Figure~\ref{fig:l63whybrid100}, the hybrid neural-physics model integrated into the DEnKF algorithm is able to accurately estimate the full state of the weakly nonlinear Lorenz system from time $t=0$ to $t=6$ even with the large time interval between two observations. \textcolor{rev3}{The largest Lyapunov exponent for the weakly nonlinear Lorenz system is 0.86 and the system eventually becomes chaotic if we integrate it for a long time. To demonstrate the capability of the proposed framework, we integrate the weakly nonlinear Lorenz system for a long lead time up to $t=30$. The LSTM network is trained using the true trajectory up to $t=15$. From Figure~\ref{fig:l63whybrid100_long}, we can see that the LSTM model diverges from the true trajectory after approximately 15 Lyapunov times. If we assimilate the observations through the DEnKF algorithm, then the analysis state is close to the true state trajectory.}

\begin{figure}[h]
\centerline{\includegraphics[width=0.98\linewidth]{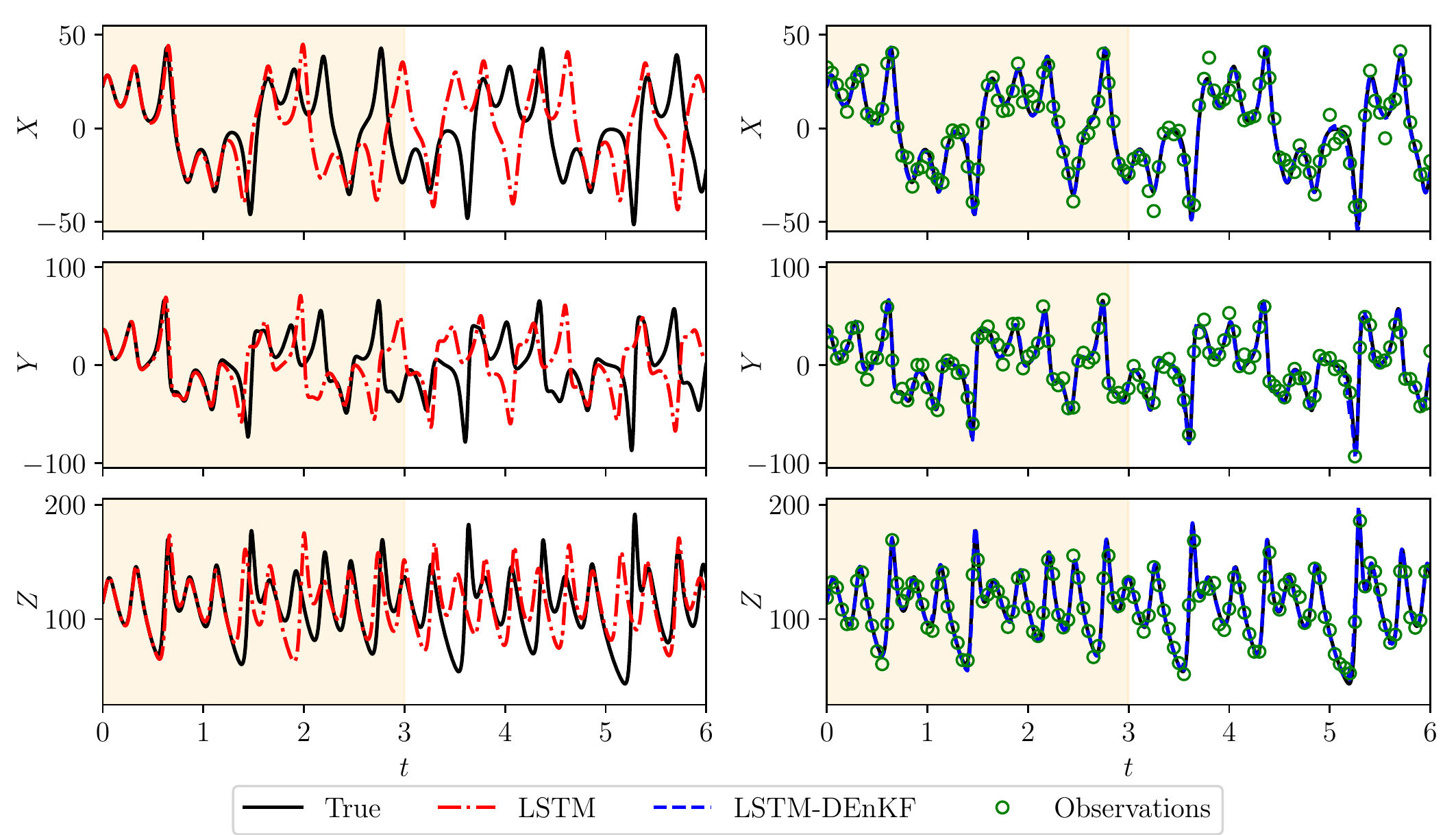}}
 \caption{The trajectory of the strongly nonlinear Lorenz system computed using the true physics-based model, the hybrid model where the $dZ/dt$ is approximated using the neural network, and the assimilated results with the hybrid model. The observation noise is drawn from $\mathcal{N}(0,5)$, and are collected after every 50 time steps, i.e., the time interval between two observations is 0.05. \textcolor{rev3}{Here, LSTM refers to forecast with the hybrid model and LSTM-DEnKF refers to an analysis state of the system.}}\label{fig:l63shybrid50}
\end{figure}

\begin{figure}[h]
\centerline{\includegraphics[width=0.98\linewidth]{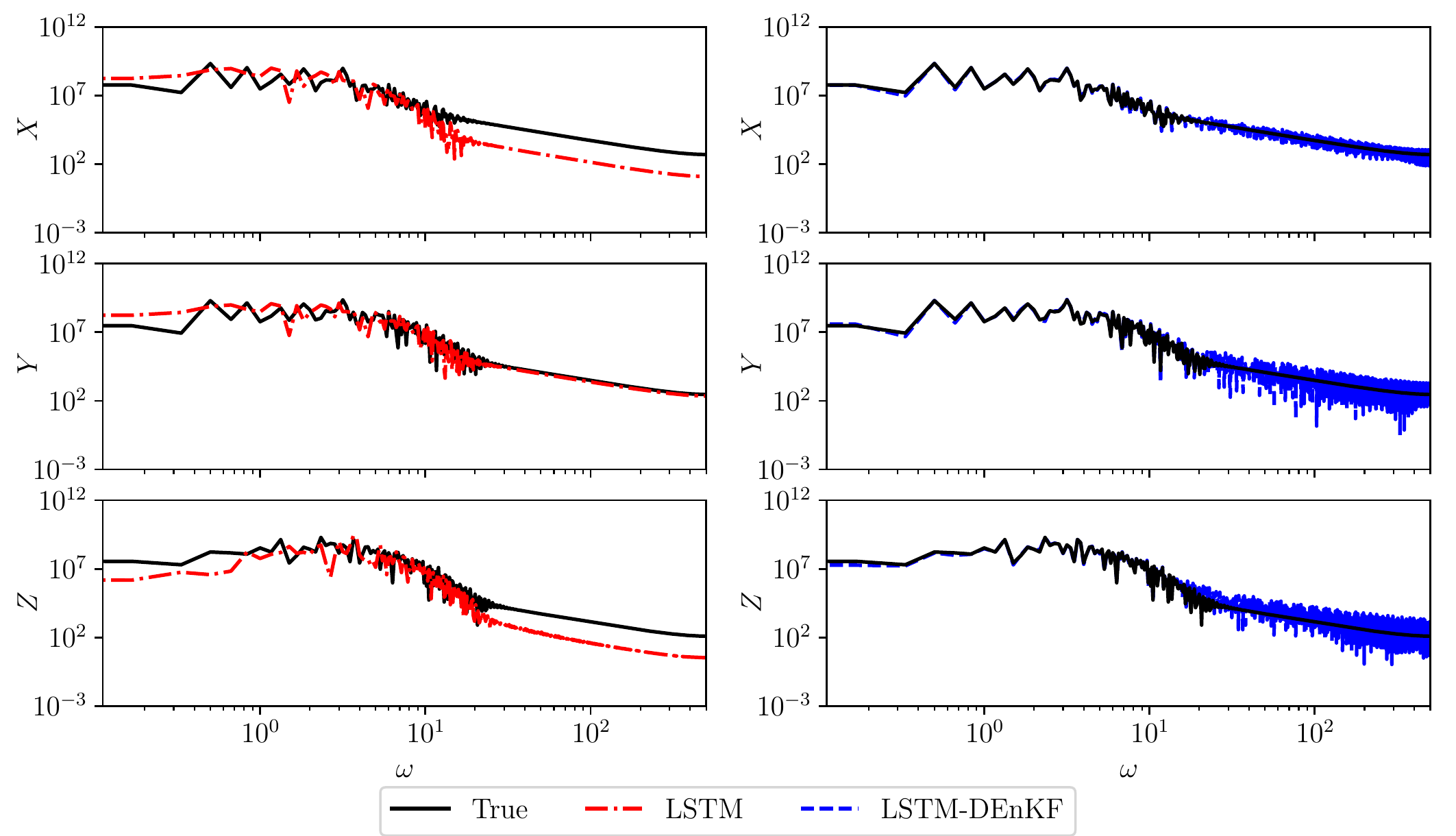}}
 \caption{Power spectra analysis of the strongly nonlinear Lorenz system computed using the true physics-based model, the hybrid model where the $dZ/dt$ is approximated using the neural network, and the assimilated results with the hybrid model. The observation noise is drawn from $\mathcal{N}(0,5)$, and are collected after every 50 time steps, i.e., the time interval between two observations is 0.05. \textcolor{rev3}{Here, LSTM refers to forecast with the hybrid model and LSTM-DEnKF refers to an analysis state of the system.}}\label{fig:l63psds}
\end{figure}

After a successful demonstration of the DEnKF algorithm with the hybrid neural-physics model for the weakly nonlinear Lorenz system, we discuss the results for the highly nonlinear Lorenz system. Figure~\ref{fig:l63shybrid50} depicts the state trajectory of the highly nonlinear Lorenz system for the true model, hybrid neural-physics model, and the hybrid neural-physics model coupled with the DEnKF. The observations for the highly nonlinear Lorenz system are contaminated using the Gaussian noise with variance $\sigma_b^2=5$ and are assimilated after every 50 time steps. Even though the LSTM is trained for time $t=0$ to $t=3$ the state predicted by the hybrid neural-physics model starts deviating from the true model trajectory at around $t \approx 1.5$. This is due to the accumulation of prediction error in the deployment of the trained LSTM network and high sensitivity to the initial condition in the strongly nonlinear Lorenz system. When the hybrid neural-physics model is coupled with the DEnKF algorithm, the predicted state is accurate beyond the training period up to time $t=6$. This shows the success of the LSTM network to learn the mapping between resolved dynamics and the missing physics of the Lorenz system and also testifies the functioning of sequential DA. Figure~\ref{fig:l63psds} shows the PSD for all three states of the highly nonlinear Lorenz system for the true model, hybrid neural-physics model, and hybrid neural-physics model coupled with DEnKF. We can observe that there is a mismatch in the distribution of PSD at lower frequencies for the state predicted by the hybrid neural-physics model compared to the true model state trajectory. The PSD for the state trajectory predicted with the hybrid neural-physics model coupled with DEnKF agrees very well with the PSD of the true model state trajectory.

\begin{figure}[h]
\centerline{\includegraphics[width=0.98\linewidth]{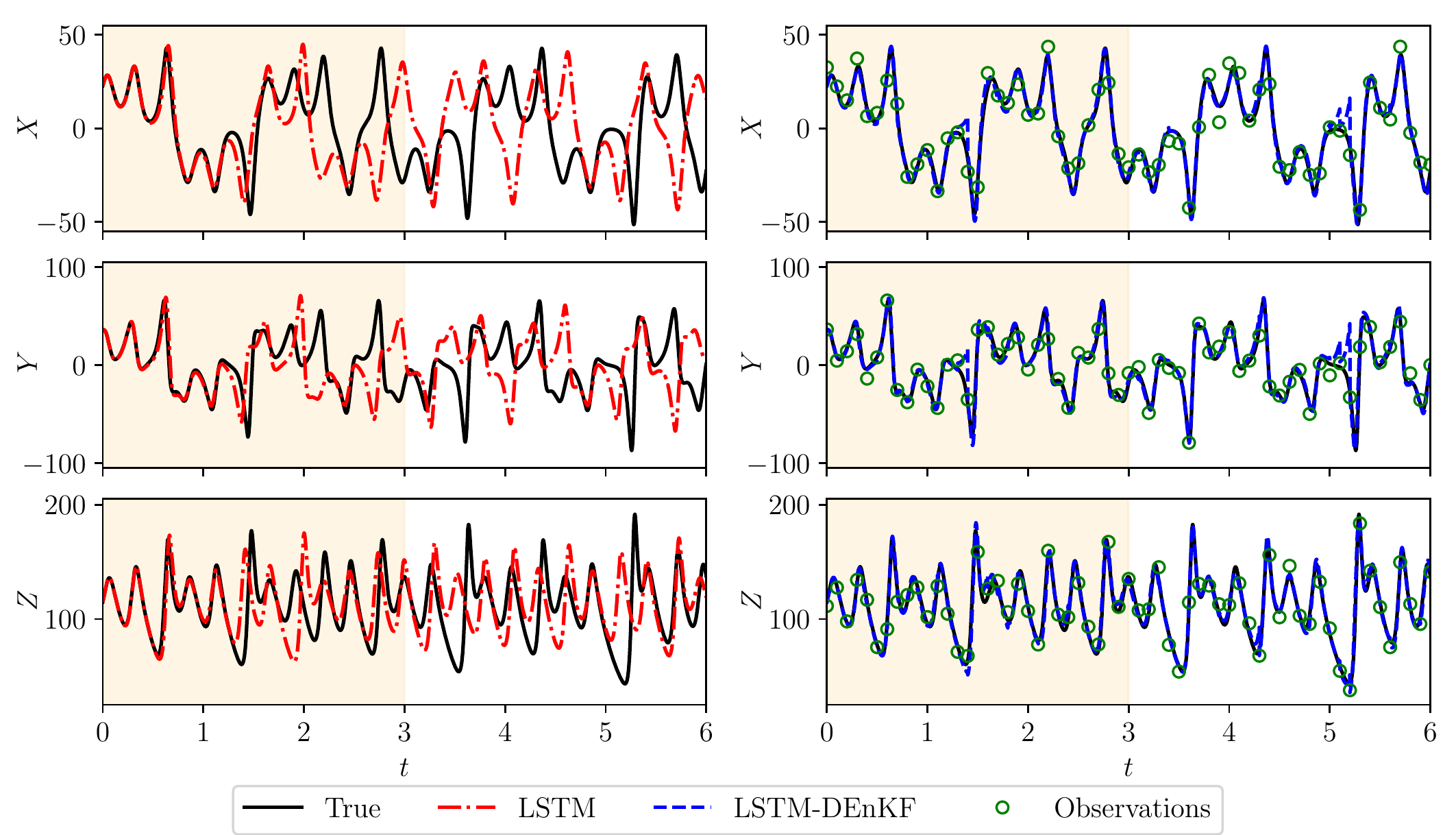}}
 \caption{The trajectory of the strongly nonlinear Lorenz system computed using the true physics-based model, the hybrid model where the $dZ/dt$ is approximated using the neural network, and the assimilated results with the hybrid model. The observation noise is drawn from $\mathcal{N}(0,5)$, and are collected after every 100 time steps, i.e., the time interval between two observations is 0.1. \textcolor{rev3}{Here, LSTM refers to forecast with the hybrid model and LSTM-DEnKF refers to an analysis state of the system.}}\label{fig:l63shybrid100}
\end{figure}

\begin{figure}[h]
\centerline{\includegraphics[width=0.98\linewidth]{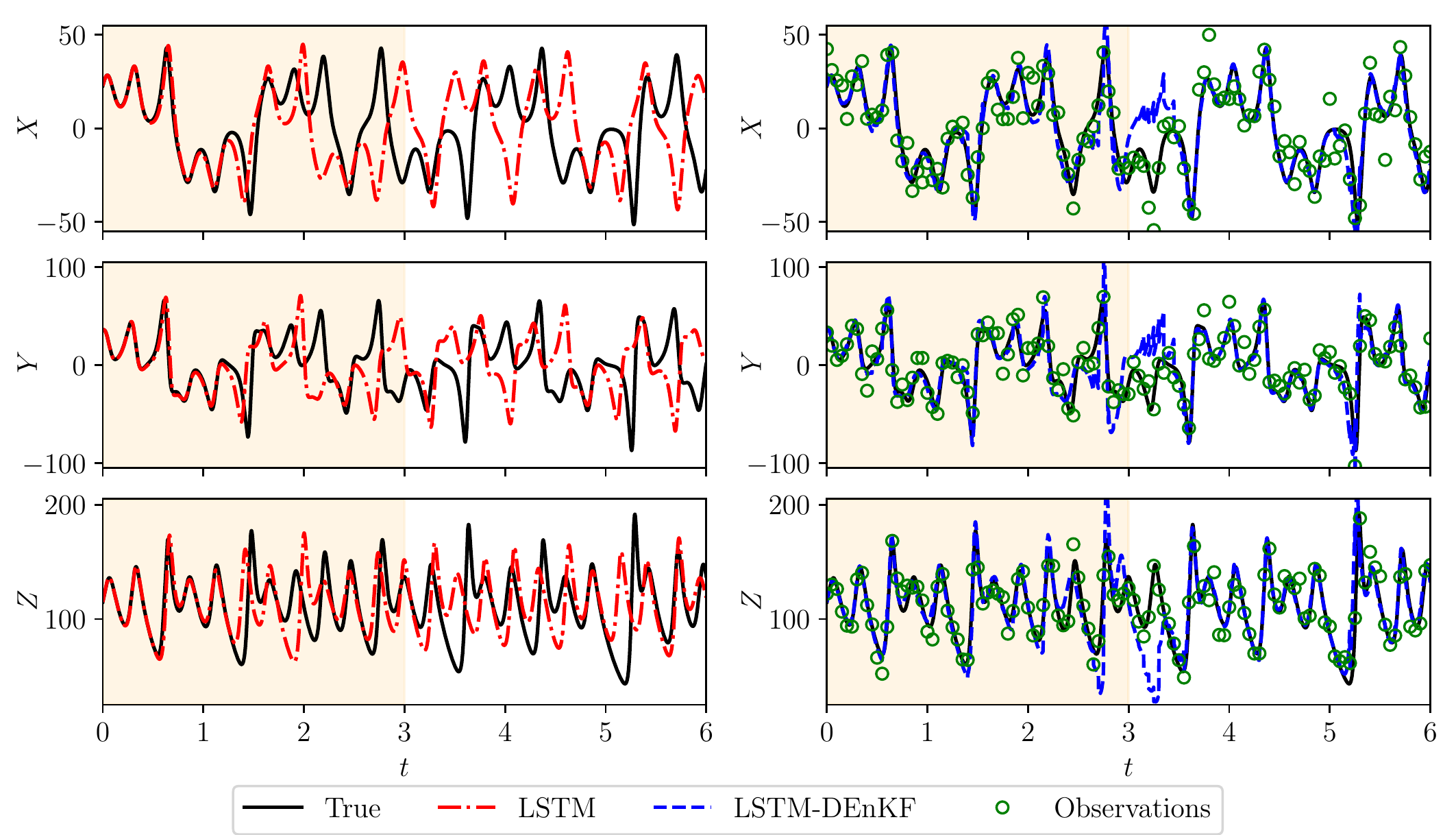}}
 \caption{The trajectory of the strongly nonlinear Lorenz system computed using the true physics-based model, the hybrid model where the $dZ/dt$ is approximated using the neural network, and the assimilated results with the hybrid model \textcolor{rev1}{using an ensemble of 10 members}. The observation noise is drawn from $\mathcal{N}(0,10)$, and are collected after every 50 time steps, i.e., the time interval between two observations is 0.05. \textcolor{rev3}{Here, LSTM refers to forecast with the hybrid model and LSTM-DEnKF refers to an analysis state of the system.}}\label{fig:l63s10hybrid100}
\end{figure}

\begin{figure}[h]
\centerline{\includegraphics[width=0.98\linewidth]{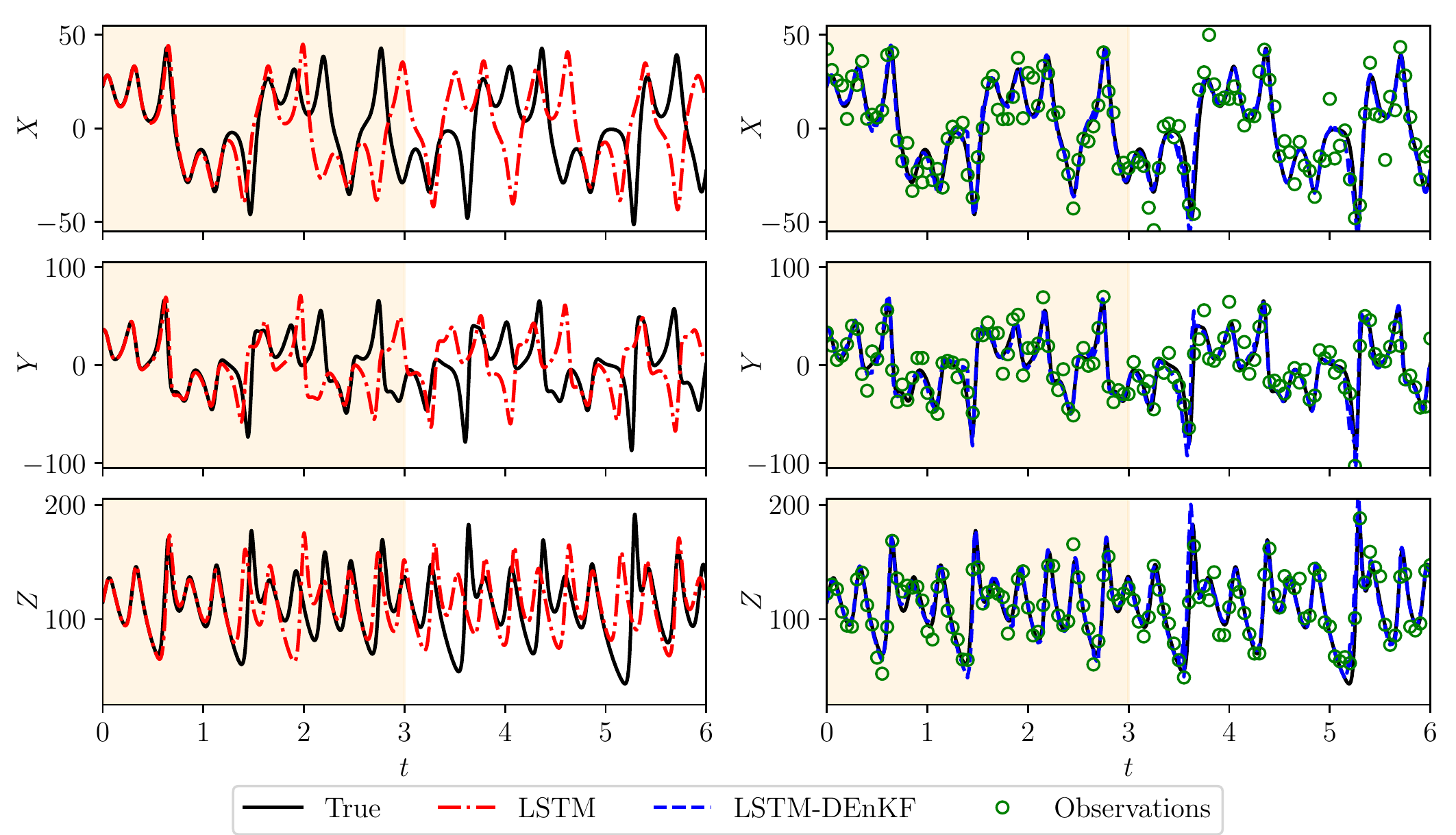}}
 \caption{The trajectory of the strongly nonlinear Lorenz system computed using the true physics-based model, the hybrid model where the $dZ/dt$ is approximated using the neural network, and the assimilated results with the hybrid model \textcolor{rev1}{using an ensemble of 20 members}. The observation noise is drawn from $\mathcal{N}(0,10)$, and are collected after every 50 time steps, i.e., the time interval between two observations is 0.05. \textcolor{rev3}{Here, LSTM refers to forecast with the hybrid model and LSTM-DEnKF refers to an analysis state of the system.}}\label{fig:l63s10hybrid100_20}
\end{figure}

In the next set of numerical experiments, we study the effect of observation noise and the frequency of assimilation on the prediction of the state of the highly nonlinear Lorenz system. Figure~\ref{fig:l63shybrid100} shows the prediction of the state trajectory of the highly nonlinear Lorenz system with observations assimilated after every 100 time steps. The observation noise is taken to be from the Gaussian distribution with variance $\sigma_b^2=5$ and the number of ensembles is set at 10. We get a very good prediction between the true state of the Lorenz system and the state predicted by the hybrid neural-physics model integrated with the DEnKF. We further increase the observation noise to the Gaussian distribution with the variance of $\sigma_b^2=10$ and assimilate observations every 50 time steps. We note here that the number of ensembles is kept fixed at 10. As shown in Figure~\ref{fig:l63s10hybrid100}, there is a discrepancy between the analyzed state estimate of the Lorenz system and the true state for the highly nonlinear case. This discrepancy can be due to the small number of ensembles in the DEnKF algorithm. In the DEnKF algorithm, the error covariance matrix is approximated by computing the statistics of all ensemble samples and ensemble size of 10 might not be enough for the large measurement noise. We repeat the numerical experiment with the same set of parameters (i.e., the observation noise and the frequency of observations) and increase the number of ensembles to 20. Figure~\ref{fig:l63s10hybrid100_20} displays the analyzed state estimate \textcolor{rev1}{using an ensemble of 20 members} for the hybrid neural-physics model coupled with DEnKF. The analyzed state estimate is improved and there is a very good agreement between the true state of the highly nonlinear Lorenz system and the analyzed state estimate by the hybrid neural physics model coupled with DEnKF. 

These findings illustrate the robustness of sequential DA in handling sparse and noisy observations to improve the state estimate of the hybrid neural-physics model. The underestimation of the error covariance matrix can also be handled using practical DA techniques like covariance inflation \cite{anderson1999monte,grudzien2018chaotic} and covariance localization \cite{hamill2001distance,kepert2009covariance}. These techniques also prevent the filter divergence when the number of ensembles is less \cite{houtekamer1998data,whitaker2012evaluating}. Indeed for high-dimensional geophysical systems, increasing the number of ensembles will be computationally expensive and adaptive inflation and localization techniques can be adopted \cite{anderson2007adaptive,kirchgessner2014choice,attia2018optimal}. \textcolor{rev1}{The quantitative performance of all numerical experiments with sequential data assimilation for the hybrid neural-physics Lorenz model is evaluated using the root mean squared error (RMSE) defined as} 
\textcolor{rev1}{
\begin{equation}\label{eq:rmse}
 \text{RMSE} = \sqrt{\frac{1}{n}\frac{1}{n_t}\sum_{i=1}^{n} \sum_{k=1}^{n_t} \big(x_i^{\text{T}}(t_k) - {x}_i^{\text{P}}(t_k)\big)^2 },
\end{equation}
where $x_i^T$ is the true state of the system and $x_i^P$ is the predicted state of the system. Table~\ref{tab:rmse} reports the RMSE for all numerical experiments carried out with a hybrid neural-physics model coupled with DA. We can notice that an increase in the time between two observations leads to poor prediction for both weak and strongly nonlinear Lorenz systems. We also observe that for the strongly nonlinear Lorenz system, the increase in observation noise deteriorates the prediction for an ensemble of 10 members. When we increase the number of ensemble members to 20, the prediction is improved and the RMSE is decreased. 
}
\begin{table*}
    \centering
    \begin{tabular}{p{0.1\textwidth}p{0.1\textwidth}p{0.1\textwidth}p{0.1\textwidth}p{0.1\textwidth}}
    \hline\noalign{\smallskip}
     Type & $\sigma_b^2$ & $N_s$ & $N$ & RMSE  \\
     \hline\noalign{\smallskip}
     Weak & 1.0 & 50 & 10 & 0.169  \\
     Weak & 1.0 & 100 & 10 & 0.233  \\
     Strong & 5.0 & 50 & 10 & 2.417  \\
     Strong & 5.0 & 100 & 10 & 4.804  \\
     Strong & 10.0 & 50 & 10 & 15.385  \\
     Strong & 10.0 & 50 & 20 & 6.889  \\
    \hline
    \end{tabular}
    \caption{Quantitative assessment of different numerical experiments carried out with hybrid neural-physics model coupled with DA. The root mean square error (RMSE) is computed using Equation~(\ref{eq:rmse}). Here, $\sigma_b^2$ is the observation noise, $N_s$ is the number of time steps between two observations and $N$ is the number of ensembles used in DEnKF algorithm.} \vspace{-5pt}
    \label{tab:rmse}
\end{table*}

\subsection{\textcolor{rev1}{Two-scale Lorenz 96 model}}

\textcolor{rev1}{The ground truth data for learning the subgrid scale parameterization in a two-scale Lorenz system is generated by temporally integrating the model using the fourth-order Runge-Kutta (RK4) time-stepping scheme with the time step $\Delta t = 0.001$ MTU. The equilibrium initial condition for the slow variables is set as $X_n=0$ for $n \in 2,\dots,N$ and $X_1=1.0$. In a similar manner, the fast variables are assigned $Y_1=1.0$ and $Y_m=0$ for $m \in 2, \dots, MN$ as an initial condition. The initial transient period of 5 MTU is disregarded. The periodic boundary condition is applied for both slow and fast variables in the two-scale Lorenz model. The output for the first 10 MTU is utilized for training and the performance of the trained network is evaluated for the next 10 MTU. Since the training and forecast period are different, there is no overlap between the train and test data. The trained LSTM network predicts the subgrid-scale parameterization at $k$th time step, i.e., $\mathbf{G}^{(k)}$ using the temporal history of resolved flow variables for six consecutive past time steps. We highlight here that we utilize the same time step for the forecast as the ground truth data generation. However, one can also use the different time step size for the forecast period by using the forecast time step size in Equation~\ref{eq:two_scale_g} to compute the parameterization term.}

\begin{figure}[h]
\centerline{\includegraphics[width=0.98\linewidth]{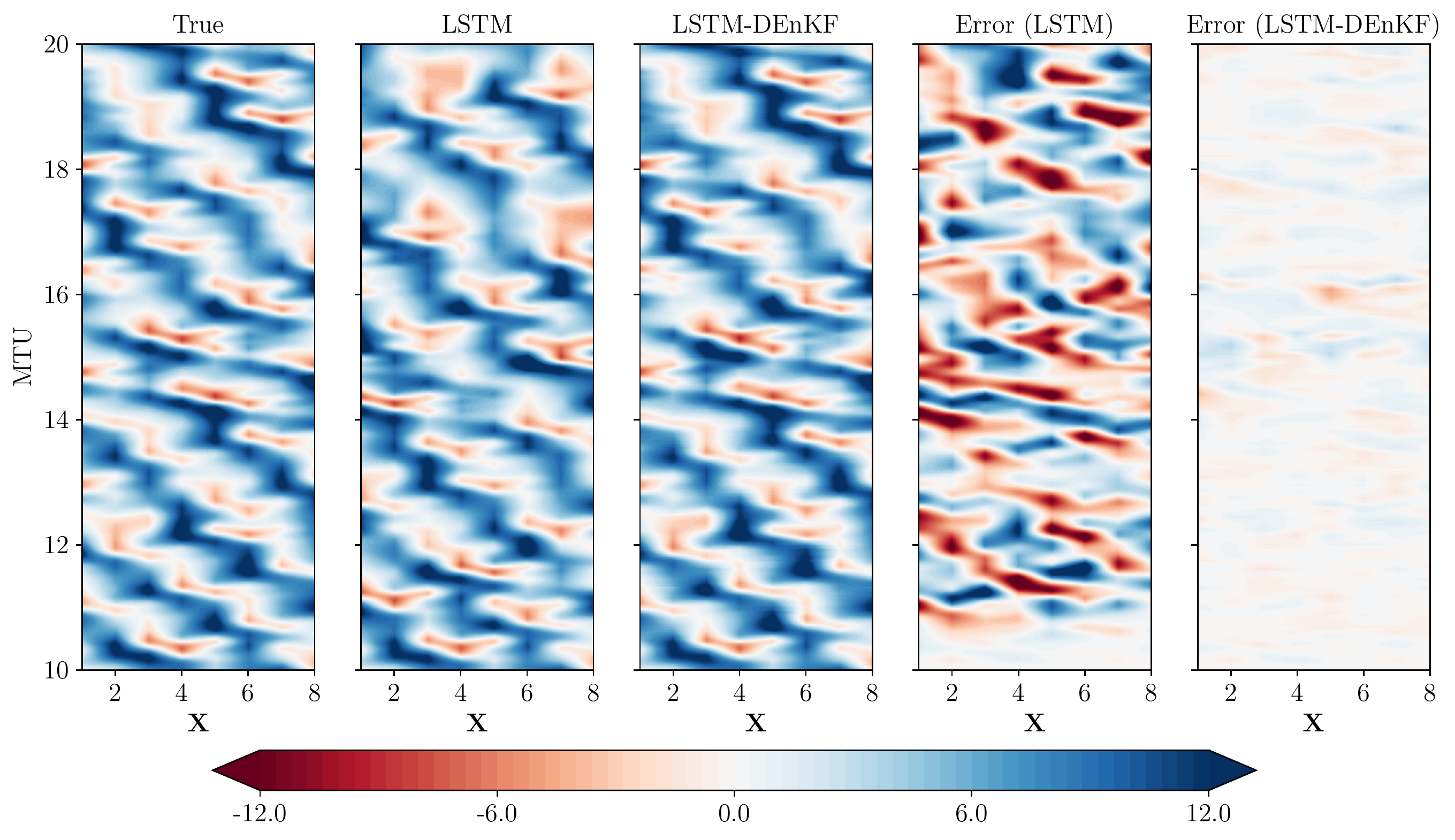}}
 \caption{Full state trajectory of the two-scale Lorenz 96 model with the closure term computed using the LSTM neural network and the DEnKF used for data assimilation. \textcolor{rev3}{Here, LSTM refers to forecast with the hybrid model and LSTM-DEnKF refers to an analysis state of the system.}}\label{fig:l2s_contour}
\end{figure}

\begin{figure}[h]
\centerline{\includegraphics[width=0.98\linewidth]{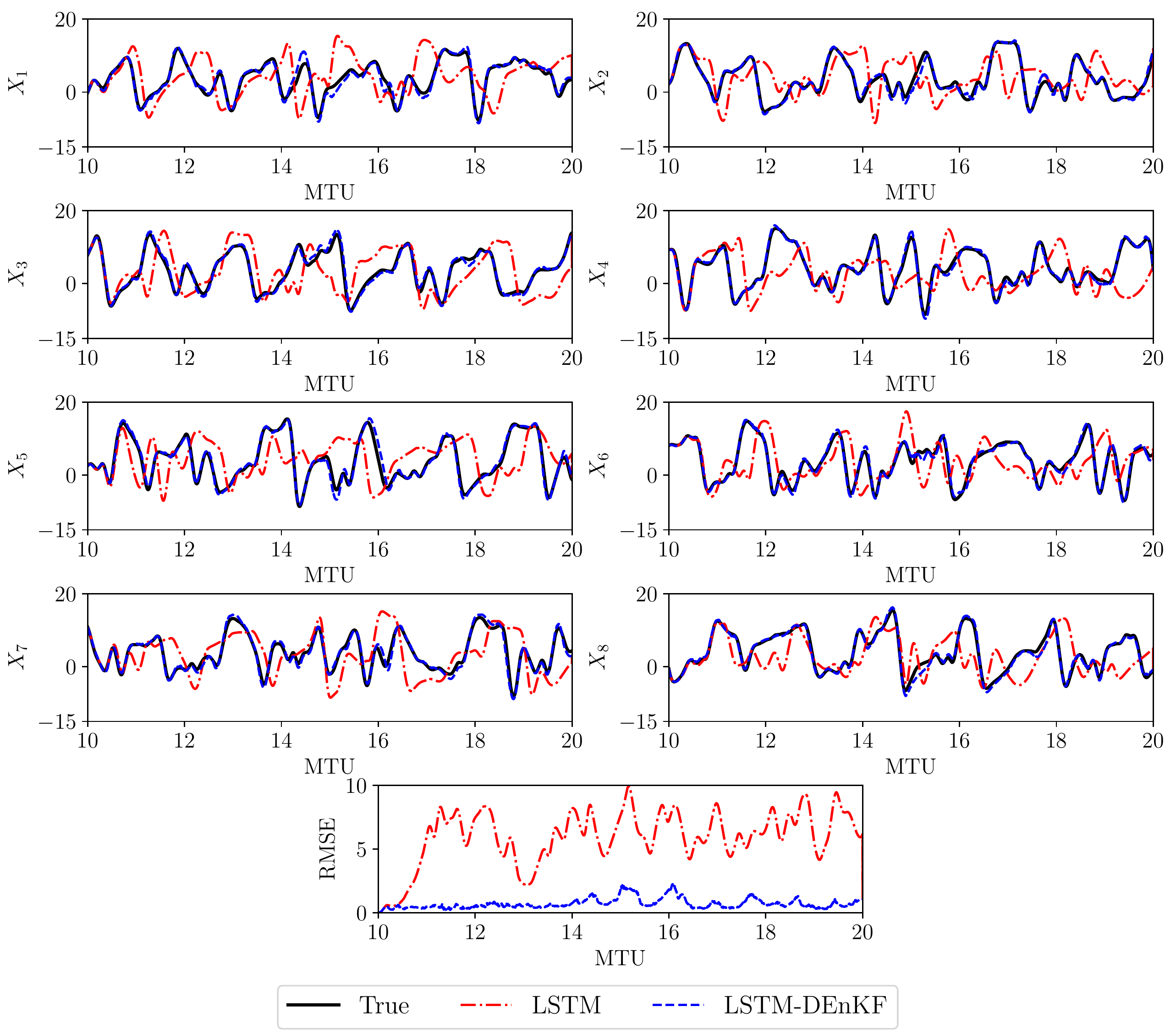}}
 \caption{The trajectory of the few selected states from the two-scale Lorenz 96 model. The bottom plot shows the evolution RMSE with time. The RMSE reported is the average mean squared error for all slow variables, i.e., $\text{RMSE}(t_k) = \sqrt{(1/N)\sum_{n=1}^{N} \big(x_n^{\text{T}}(t_k) - {x}_n^{\text{P}}(t_k)\big)^2 }$. \textcolor{rev3}{Here, LSTM refers to forecast with the hybrid model and LSTM-DEnKF refers to an analysis state of the system.}}\label{fig:l2s_ts}
\end{figure}

\textcolor{rev1}{During the online deployment, we also use the DEnKF algorithm to improve the state prediction. The observation data for the DEnKF algorithm are generated by adding the Gaussian noise with the variance $\sigma_b^2=1.0$ and the observations are assimilated every 10 time steps, i.e., 0.01 MTU. We assume that 50\% of the full state is observable and the observation points correspond to $X_1, X_3, X_5,$ and $X_7$. We initialize 10 different ensemble members by adding the Gaussian noise with the variance $\sigma_0^2=1 \times 10^{-2}$ to the true initial state of the system. We also apply an inflation factor $\lambda=1.02$ to account for modeling errors. Figure~\ref{fig:l2s_contour} displays the full-state trajectory of the slow variables in the two-scale Lorenz system from 10 MTU to 20 MTU (forecast period) along with the difference between the true and predicted state. The true state is calculated by solving both the evolution of slow and fast variables. We see that the LSTM based parameterization provides accurate prediction only for approximately 1 MTU and then the prediction has diverged from the true state of the two-scale Lorenz system. We also notice that the LSTM based subgrid scale parameterization model coupled with DEnKF renders accurate state prediction for the entire period of the forecast. In Figure~\ref{fig:l2s_ts}, the trajectory of each slow variable is shown and we see that the LSTM-DEnKF framework can estimate the state for all slow variables with very high accuracy. The temporal variation of the RMSE also suggests that the sequential data assimilation aids in considerable improvement in the state prediction compared to employing only the trained LSTM network.}

\begin{figure}[h]
\centerline{\includegraphics[width=0.98\linewidth]{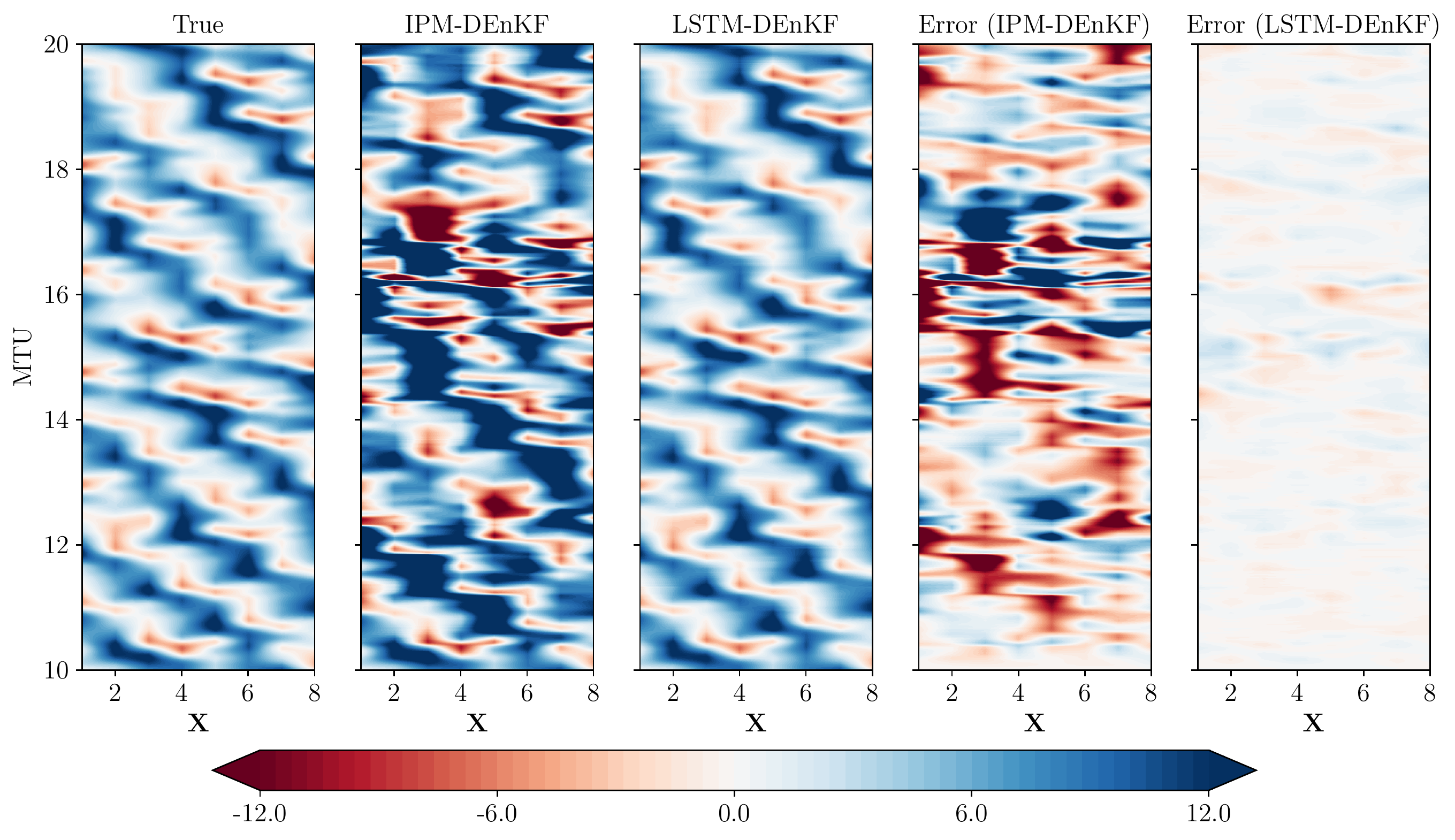}}
 \caption{\textcolor{rev3}{Full state trajectory of the two-scale Lorenz 96 model with incomplete physics-based model (IPM) and the closure term computed using the LSTM neural network. The forecast for both models is further corrected using  DEnKF applied for data assimilation.}}\label{fig:l2s_contour_ipm}
\end{figure}

\textcolor{rev3}{We highlight here that it may be argued that the forecast of a free run and an analysis (corrected forecast from DEnKF) of the system do not have access to the same information. Specifically, the LSTM-DEnKF model utilizes sparse and noisy observations that are not available to the LSTM model. For a fair comparison, we repeat the numerical experiments with the incomplete physics-based model (IPM) coupled with the DEnKF model. Figure~\ref{fig:l2s_contour_ipm} depicts the full state trajectory of the slow variables in the two-scale Lorenz system from 10 MTU to 20 MTU along with the difference between the true and predicted state. We can see that the analysis with the LSTM model is significantly more accurate than the analysis with the IPM model. This result demonstrates that machine learning offers possibilities of learning missing physics or model imperfections and the data assimilation prediction with such hybrid models is superior compared to truncated models.}

\begin{figure}[h]
\centerline{\includegraphics[width=0.95\linewidth]{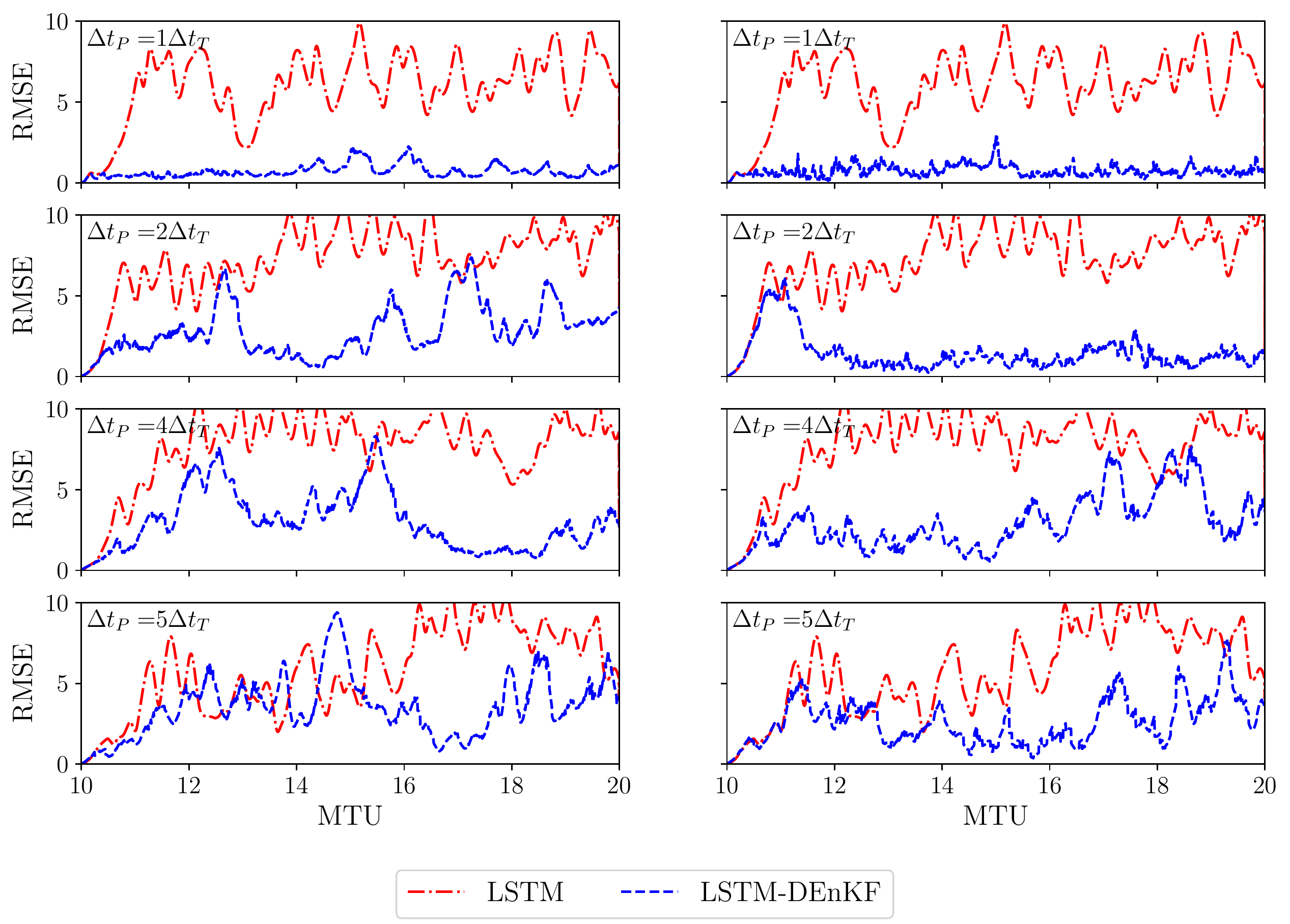}}
 \caption{\textcolor{rev2}{Evolution of the RMSE for the two-scale Lorenz 96 model with different time step used in the prediction for $\sigma_b^2=1.0$ (left) and $\sigma_b^2=10.0$ (right). The $\Delta t_T$ represents the time step used for generating the training data, and $\Delta t_P$ is the time step used during prediction. The reported RMSE is the average mean squared error for all slow variables, i.e., $\text{RMSE}(t_k) = \sqrt{(1/N)\sum_{n=1}^{N} \big(x_n^{\text{T}}(t_k) - {x}_n^{\text{P}}(t_k)\big)^2 }$.} \textcolor{rev3}{Here, LSTM refers to forecast with the hybrid model and LSTM-DEnKF refers to an analysis state of the system.}}\label{fig:dt_sensitivity}
\end{figure}

\textcolor{rev2}{As discussed in Section~\ref{sec:l2s}, the tendency (i.e., the time derivative) is approximated by a finite-difference method in the computation of subgrid scale parameterization. One of the important questions is the effect of this approximation on the generalizability of the trained neural network. We investigate this effect by deploying the trained neural network in a two-scale Lorenz 96 system with a different time step used during the deployment. A similar sensitivity study was also conducted in Brajard et al. \cite{brajard2020combining}, where they analyzed the effect of time step on the linear superposition assumption between the resolved and unresolved part of the model. Figure~\ref{fig:dt_sensitivity} shows the evolution of the RMSE for the two-scale Lorenz 96 model with a different time step used in the prediction and for different observation noise. \textcolor{rev3}{In Figure~\ref{fig:dt_sensitivity}, the $\Delta t_T$ denotes the time step used for generating the training data, and $\Delta t_P$ is the time step used during prediction. For example, the LSTM network is trained using the data generated with a time step of 0.001 MTU, and $\Delta t_P=2\Delta t_T$ means that the time step during the forecast stage is 0.002 MTU. For all numerical experiments with a two-scale Lorenz 96 system, the observations are assimilated every 10 prediction time steps ( i.e., $\Delta t_P$).} We can observe that the RMSE is slightly increased for the LSTM based subgrid scale parameterization model with a different time step utilized during prediction. We do not observe any unstable behavior with a different time step for prediction than the time step used for training. The RMSE for the LSTM-DEnKF model is higher when there is a difference between the prediction and training time step. However, there is still a considerable improvement in the prediction compared to employing only LSTM based subgrid scale parameterization. The numerical experiment with a large observation noise, i.e., $\sigma_b^2=10.0$ suggests that the framework is robust for high uncertainty in measurements.}

\section{Summary and conclusions} \label{sec:conclusion}
In this work, we investigated the use of a neural network to recover the missing physics (i.e., incomplete dynamics) in dynamical systems. The hybrid neural-physics model composed of the physics-based model for the known dynamics and neural network for the unknown dynamics is applied to the Lorenz system with different strengths of nonlinearity. In the Lorenz system, the third equation was assumed to be missing and was approximated using the neural network. The recurrent neural network capable of handling long and short term dependencies was utilized for learning the mapping between known and unknown dynamics. Once the trained long short-term memory (LSTM) network is deployed in an auto-regressive manner, there is an accumulation of prediction error from one time to another. The weakly nonlinear Lorenz system exhibits near-regular oscillations with increasing amplitude and the hybrid neural-physics model is able to produce an accurate forecast even in the presence of prediction error during the online deployment of the LSTM network. The highly nonlinear Lorenz system being chaotic with extreme events is sensitive to the initial condition and a small change in the initial condition can produce a very different output. We observe that the predicted state trajectory of the hybrid neural-physics model starts deviating from the true trajectory after around 3.5 multiple of Lyapunov time for the highly nonlinear Lorenz system. 

To achieve an accurate prediction over a longer period, the sequential data assimilation based on the deterministic ensemble Kalman filter (DEnKF) is applied. The hybrid neural-physics model coupled with DEnKF is able to predict the state of the highly nonlinear Lorenz system close to the true state by utilizing sparse and noisy observations. The analysis with the power spectral density (PSD) demonstrates that the hybrid neural-physics model integrated with DEnKF can capture the power density at lower frequencies accurately compared with the true PSD. \textcolor{rev1}{We also illustrate the successful performance of the present framework for the two-scale Lorenz system where the subgrid scale parameterization is learned through the LSTM network. During the forecast period, the DEnKF algorithm is utilized to make use of sparse and noisy observations to improve the prediction.} \textcolor{rev3}{The DEnKF algorithm fail to give accurate analysis if unresolved scales are not parameterized (i.e., incomplete physics-based model).} \textcolor{rev2}{Furthermore, we show that the hybrid model forecast is not highly sensitive to the time step used for computing the subgrid-scale tendencies and the framework is robust against high observational noise.} Based on numerical experiments carried out in this work, the methodology presented here seems promising for a continuous forecast of incomplete chaotic dynamical systems. 

It may be argued that the success of this framework relies on the low-dimensionality of the Lorenz model. \textcolor{rev1}{Even the two-scale Lorenz model has many fewer dimensions than any complex geophysical system and a relatively simple parameterization.} Nevertheless, we foresee that the approach can be extended to high-dimensional complex systems by considering advanced neural network architectures like convolutional LSTMs, generative adversarial networks (GANs). The proposed method is modular enough that it is possible to plug any type of neural network and data assimilation scheme independently. The lessons learned from this study will guide our future extension of this approach to more complex chaotic systems whose dynamics is representative of geophysical flows. Moreover, a relevant possible extension is to combine machine learning, data assimilation, and reduced order models toward accelerating digital transformation and decision making processes in science, technology, and engineering workflows. \textcolor{rev1}{To utilize the present framework for real-time decision making, it is necessary to reduce the computational requirement of the forward model significantly. Therefore, replacing the numerical discretization-based solver with the data-driven surrogate model will be a future research direction.}

\section*{Acknowledgements}
This material is based upon work supported by the U.S. Department of Energy, Office of Science, Office of Advanced Scientific Computing Research under Award Number DE-SC0019290. O.S. gratefully acknowledges their support. Disclaimer: This report was prepared as an account of work sponsored by an agency of the United States Government. Neither the United States Government nor any agency thereof, nor any of their employees, makes any warranty, express or implied, or assumes any legal liability or responsibility for the accuracy, completeness, or usefulness of any information, apparatus, product, or process disclosed, or represents that its use would not infringe privately owned rights. Reference herein to any specific commercial product, process, or service by trade name, trademark, manufacturer, or otherwise does not necessarily constitute or imply its endorsement, recommendation, or favoring by the United States Government or any agency thereof. The views and opinions of authors expressed herein do not necessarily state or reflect those of the United States Government or any agency thereof.

\bibliographystyle{unsrt} 
\bibliography{ref}

\begin{thebibliography}{10}

\bibitem{schneider2017earth}
Tapio Schneider, Shiwei Lan, Andrew Stuart, and Joao Teixeira.
\newblock Earth system modeling 2.0: A blueprint for models that learn from
  observations and targeted high-resolution simulations.
\newblock {\em Geophysical Research Letters}, 44(24):12--396, 2017.

\bibitem{wood2012stratocumulus}
Robert Wood.
\newblock Stratocumulus clouds.
\newblock {\em Monthly Weather Review}, 140(8):2373--2423, 2012.

\bibitem{fox2014principles}
B~Fox-Kemper, S~Bachman, B~Pearson, and S~Reckinger.
\newblock Principles and advances in subgrid modelling for eddy-rich
  simulations.
\newblock {\em Clivar Exchanges}, 19(2):42--46, 2014.

\bibitem{schneider2017climate}
Tapio Schneider, Jo{\~a}o Teixeira, Christopher~S Bretherton, Florent Brient,
  Kyle~G Pressel, Christoph Sch{\"a}r, and A~Pier Siebesma.
\newblock Climate goals and computing the future of clouds.
\newblock {\em Nature Climate Change}, 7(1):3--5, 2017.

\bibitem{barnett1993enso}
TP~Barnett, N~Graham, S~Pazan, W~White, Mojib Latif, and M~Fl{\"u}gel.
\newblock {ENSO and ENSO-related predictability. Part I: Prediction of
  equatorial Pacific sea surface temperature with a hybrid coupled
  ocean--atmosphere model}.
\newblock {\em Journal of Climate}, 6(8):1545--1566, 1993.

\bibitem{hornik1991approximation}
Kurt Hornik.
\newblock Approximation capabilities of multilayer feedforward networks.
\newblock {\em Neural networks}, 4(2):251--257, 1991.

\bibitem{lee1997application}
Changhoon Lee, John Kim, David Babcock, and Rodney Goodman.
\newblock Application of neural networks to turbulence control for drag
  reduction.
\newblock {\em Physics of Fluids}, 9(6):1740--1747, 1997.

\bibitem{tang2001neural}
Youmin Tang, William~W Hsieh, Benyang Tang, and Keith Haines.
\newblock A neural network atmospheric model for hybrid coupled modelling.
\newblock {\em Climate Dynamics}, 17(5-6):445--455, 2001.

\bibitem{tang2001coupling}
Youmin Tang and William~W Hsieh.
\newblock Coupling neural networks to incomplete dynamical systems via
  variational data assimilation.
\newblock {\em Monthly Weather Review}, 129(4):818--834, 2001.

\bibitem{liaqat2003applying}
Ali Liaqat, Makoto Fukuhara, and Tatsuoki Takeda.
\newblock Applying a neural network collocation method to an incompletely known
  dynamical system via weak constraint data assimilation.
\newblock {\em Monthly Weather Review}, 131(8):1696--1714, 2003.

\bibitem{krasnopolsky2005new}
Vladimir~M Krasnopolsky, Michael~S Fox-Rabinovitz, and Dmitry~V Chalikov.
\newblock New approach to calculation of atmospheric model physics: Accurate
  and fast neural network emulation of longwave radiation in a climate model.
\newblock {\em Monthly Weather Review}, 133(5):1370--1383, 2005.

\bibitem{sutskever2014sequence}
Ilya Sutskever, Oriol Vinyals, and Quoc~V Le.
\newblock Sequence to sequence learning with neural networks.
\newblock In {\em Advances in neural information processing systems}, pages
  3104--3112, 2014.

\bibitem{krizhevsky2017imagenet}
Alex Krizhevsky, Ilya Sutskever, and Geoffrey~E Hinton.
\newblock Imagenet classification with deep convolutional neural networks.
\newblock {\em Communications of the ACM}, 60(6):84--90, 2017.

\bibitem{mnih2015human}
Volodymyr Mnih, Koray Kavukcuoglu, David Silver, Andrei~A Rusu, Joel Veness,
  Marc~G Bellemare, Alex Graves, Martin Riedmiller, Andreas~K Fidjeland, Georg
  Ostrovski, et~al.
\newblock Human-level control through deep reinforcement learning.
\newblock {\em Nature}, 518(7540):529--533, 2015.

\bibitem{goodfellow2014generative}
Ian Goodfellow, Jean Pouget-Abadie, Mehdi Mirza, Bing Xu, David Warde-Farley,
  Sherjil Ozair, Aaron Courville, and Yoshua Bengio.
\newblock Generative adversarial nets.
\newblock In {\em Advances in neural information processing systems}, pages
  2672--2680, 2014.

\bibitem{reichstein2019deep}
Markus Reichstein, Gustau Camps-Valls, Bjorn Stevens, Martin Jung, Joachim
  Denzler, Nuno Carvalhais, et~al.
\newblock {Deep learning and process understanding for data-driven Earth system
  science}.
\newblock {\em Nature}, 566(7743):195--204, 2019.

\bibitem{mcgovern2017using}
Amy McGovern, Kimberly~L Elmore, David~John Gagne, Sue~Ellen Haupt,
  Christopher~D Karstens, Ryan Lagerquist, Travis Smith, and John~K Williams.
\newblock Using artificial intelligence to improve real-time decision-making
  for high-impact weather.
\newblock {\em Bulletin of the American Meteorological Society},
  98(10):2073--2090, 2017.

\bibitem{sonnewald2021bridging}
Maike Sonnewald, Redouane Lguensat, Daniel~C. Jones, Peter~D. Dueben, Julien
  Brajard, and Venkatramani Balaji.
\newblock Bridging observation, theory and numerical simulation of the ocean
  using machine learning.
\newblock {\em arXiv: 2104.12506}, 2021.

\bibitem{von2019informed}
Laura Von~Rueden, Sebastian Mayer, Jochen Garcke, Christian Bauckhage, and
  Jannis Schuecker.
\newblock Informed machine learning--towards a taxonomy of explicit integration
  of knowledge into machine learning.
\newblock {\em Learning}, 18:19--20, 2019.

\bibitem{kashinath2021physics}
K~Kashinath, M~Mustafa, A~Albert, JL~Wu, C~Jiang, S~Esmaeilzadeh,
  K~Azizzadenesheli, R~Wang, A~Chattopadhyay, and A~Singh.
\newblock Physics-informed machine learning: case studies for weather and
  climate modelling.
\newblock {\em Philosophical Transactions of the Royal Society A},
  379(2194):20200093, 2021.

\bibitem{carrassi2018data}
Alberto Carrassi, Marc Bocquet, Laurent Bertino, and Geir Evensen.
\newblock Data assimilation in the geosciences: An overview of methods, issues,
  and perspectives.
\newblock {\em WIREs Climate Change}, 9(5):e535, 2018.

\bibitem{eyre2020assimilation}
Jonathan~Robert Eyre, Stephen~J English, and Mary Forsythe.
\newblock Assimilation of satellite data in numerical weather prediction. part
  i: The early years.
\newblock {\em Quarterly Journal of the Royal Meteorological Society},
  146(726):49--68, 2020.

\bibitem{rasp2018deep}
Stephan Rasp, Michael~S Pritchard, and Pierre Gentine.
\newblock Deep learning to represent subgrid processes in climate models.
\newblock {\em Proceedings of the National Academy of Sciences},
  115(39):9684--9689, 2018.

\bibitem{gentine2018could}
Pierre Gentine, Mike Pritchard, Stephan Rasp, Gael Reinaudi, and Galen Yacalis.
\newblock Could machine learning break the convection parameterization
  deadlock?
\newblock {\em Geophysical Research Letters}, 45(11):5742--5751, 2018.

\bibitem{gagne2020machine}
David~John Gagne, Hannah~M Christensen, Aneesh~C Subramanian, and Adam~H
  Monahan.
\newblock Machine learning for stochastic parameterization: Generative
  adversarial networks in the {Lorenz'96} model.
\newblock {\em Journal of Advances in Modeling Earth Systems},
  12(3):e2019MS001896, 2020.

\bibitem{pathak2018model}
Jaideep Pathak, Brian Hunt, Michelle Girvan, Zhixin Lu, and Edward Ott.
\newblock Model-free prediction of large spatiotemporally chaotic systems from
  data: A reservoir computing approach.
\newblock {\em Physical Review Letters}, 120(2):024102, 2018.

\bibitem{vlachas2018data}
Pantelis~R Vlachas, Wonmin Byeon, Zhong~Y Wan, Themistoklis~P Sapsis, and
  Petros Koumoutsakos.
\newblock Data-driven forecasting of high-dimensional chaotic systems with long
  short-term memory networks.
\newblock {\em Proceedings of the Royal Society A: Mathematical, Physical and
  Engineering Sciences}, 474(2213):20170844, 2018.

\bibitem{tang2020deep}
Meng Tang, Yimin Liu, and Louis~J Durlofsky.
\newblock A deep-learning-based surrogate model for data assimilation in
  dynamic subsurface flow problems.
\newblock {\em Journal of Computational Physics}, page 109456, 2020.

\bibitem{hsieh1998applying}
William~W Hsieh and Benyang Tang.
\newblock Applying neural network models to prediction and data analysis in
  meteorology and oceanography.
\newblock {\em Bulletin of the American Meteorological Society},
  79(9):1855--1870, 1998.

\bibitem{abarbanel2018machine}
Henry~DI Abarbanel, Paul~J Rozdeba, and Sasha Shirman.
\newblock Machine learning: Deepest learning as statistical data assimilation
  problems.
\newblock {\em Neural Computation}, 30(8):2025--2055, 2018.

\bibitem{san2021hybrid}
Omer San, Adil Rasheed, and Trond Kvamsdal.
\newblock Hybrid analysis and modeling, eclecticism, and multifidelity
  computing toward digital twin revolution.
\newblock {\em GAMM-Mitteilungen}, 44:e202100007, 2021.

\bibitem{bocquet2020bayesian}
Marc Bocquet, Julien Brajard, Alberto Carrassi, and Laurent Bertino.
\newblock Bayesian inference of chaotic dynamics by merging data assimilation,
  machine learning and expectation-maximization.
\newblock {\em Foundations of Data Science}, 2(1):55--80, 2020.

\bibitem{bonavita2020machine}
Massimo Bonavita and Patrick Laloyaux.
\newblock Machine learning for model error inference and correction.
\newblock {\em Journal of Advances in Modeling Earth Systems},
  12(12):e2020MS002232, 2020.

\bibitem{farchi2020using}
Alban Farchi, Patrick Laloyaux, Massimo Bonavita, and Marc Bocquet.
\newblock Using machine learning to correct model error in data assimilation
  and forecast applications.
\newblock {\em arXiv preprint arXiv:2010.12605}, 2020.

\bibitem{farchi2021comparison}
Alban Farchi, Marc Bocquet, Patrick Laloyaux, Massimo Bonavita, and Quentin
  Malartic.
\newblock A comparison of combined data assimilation and machine learning
  methods for offline and online model error correction.
\newblock {\em arXiv preprint arXiv:2107.11114}, 2021.

\bibitem{geer2021learning}
Alan~J Geer.
\newblock Learning earth system models from observations: machine learning or
  data assimilation?
\newblock {\em Philosophical Transactions of the Royal Society A},
  379(2194):20200089, 2021.

\bibitem{navon1998practical}
IM~Navon.
\newblock Practical and theoretical aspects of adjoint parameter estimation and
  identifiability in meteorology and oceanography.
\newblock {\em Dynamics of Atmospheres and Oceans}, 27(1-4):55--79, 1998.

\bibitem{Zhu98impactof}
Yanqiu Zhu and I.~M. Navon.
\newblock Impact of parameter estimation on the performance of the {FSU} global
  spectral model using its full-physics adjoint.
\newblock {\em Monthly Weather Review}, 127:1497--1517, 1998.

\bibitem{courtier1998ecmwf}
Philippe Courtier, E~Andersson, W~Heckley, D~Vasiljevic, M~Hamrud,
  A~Hollingsworth, F~Rabier, M~Fisher, and J~Pailleux.
\newblock The {ECMWF} implementation of three-dimensional variational
  assimilation ({3D-Var}). {I: Formulation}.
\newblock {\em Quarterly Journal of the Royal Meteorological Society},
  124(550):1783--1807, 1998.

\bibitem{barker2004three}
Dale~M Barker, Wei Huang, Yong-Run Guo, AJ~Bourgeois, and QN~Xiao.
\newblock A three-dimensional variational data assimilation system for {MM5}:
  Implementation and initial results.
\newblock {\em Monthly Weather Review}, 132(4):897--914, 2004.

\bibitem{brajard2020combining}
Julien Brajard, Alberto Carrassi, Marc Bocquet, and Laurent Bertino.
\newblock Combining data assimilation and machine learning to infer unresolved
  scale parametrization.
\newblock {\em Philosophical Transactions of the Royal Society A: Mathematical,
  Physical and Engineering Sciences}, 379(2194):20200086, 2021.

\bibitem{brajard2019representing}
Julien Brajard, Anastase Charantonis, and Jérôme Sirven.
\newblock Representing ill-known parts of a numerical model using a machine
  learning approach.
\newblock {\em arXiv: 1903.07358}, 2019.

\bibitem{pawar2020long}
Suraj Pawar, Shady~E Ahmed, Omer San, Adil Rasheed, and Ionel~M Navon.
\newblock Long short-term memory embedded nudging schemes for nonlinear data
  assimilation of geophysical flows.
\newblock {\em Physics of Fluids}, 32(7):076606, 2020.

\bibitem{pawar2021data}
Suraj Pawar and Omer San.
\newblock Data assimilation empowered neural network parametrizations for
  subgrid processes in geophysical flows.
\newblock {\em Physical Review Fluids}, 6(5):050501, 2021.

\bibitem{ahmed2021closures}
Shady~E Ahmed, Suraj Pawar, Omer San, Adil Rasheed, Traian Iliescu, and Bernd~R
  Noack.
\newblock On closures for reduced order models -- a spectrum of first-principle
  to machine-learned avenues.
\newblock {\em arXiv preprint arXiv:2106.14954}, 2021.

\bibitem{BRAJARD2020101171}
Julien Brajard, Alberto Carrassi, Marc Bocquet, and Laurent Bertino.
\newblock Combining data assimilation and machine learning to emulate a
  dynamical model from sparse and noisy observations: A case study with the
  {Lorenz} 96 model.
\newblock {\em Journal of Computational Science}, 44:101171, 2020.

\bibitem{GitHub}
Suraj Pawar.
\newblock {Hybrid\_Lorenz}, 2021.

\bibitem{lorenz1963deterministic}
Edward~N Lorenz.
\newblock Deterministic nonperiodic flow.
\newblock {\em Journal of the Atmospheric Sciences}, 20(2):130--141, 1963.

\bibitem{durran1991third}
Dale~R Durran.
\newblock The third-order {Adams-Bashforth} method: An attractive alternative
  to leapfrog time differencing.
\newblock {\em Monthly Weather Review}, 119(3):702--720, 1991.

\bibitem{kaplan1979preturbulence}
James~L Kaplan and James~A Yorke.
\newblock Preturbulence: a regime observed in a fluid flow model of {Lorenz}.
\newblock {\em Communications in Mathematical Physics}, 67(2):93--108, 1979.

\bibitem{gauthier1992chaos}
Pierre Gauthier.
\newblock Chaos and quadri-dimensional data assimilation: A study based on the
  {Lorenz} model.
\newblock {\em Tellus A: Dynamic Meteorology and Oceanography}, 44(1):2--17,
  1992.

\bibitem{elsner1992nonlinear}
JB~Elsner and AA~Tsonis.
\newblock Nonlinear prediction, chaos, and noise.
\newblock {\em Bulletin of the American Meteorological Society}, 73(1):49--60,
  1992.

\bibitem{williams1989learning}
Ronald~J Williams and David Zipser.
\newblock A learning algorithm for continually running fully recurrent neural
  networks.
\newblock {\em Neural Computation}, 1(2):270--280, 1989.

\bibitem{miller1994advanced}
Robert~N Miller, Michael Ghil, and Francois Gauthiez.
\newblock Advanced data assimilation in strongly nonlinear dynamical systems.
\newblock {\em Journal of the Atmospheric Sciences}, 51(8):1037--1056, 1994.

\bibitem{evensen1997advanced}
Geir Evensen.
\newblock Advanced data assimilation for strongly nonlinear dynamics.
\newblock {\em Monthly Weather Review}, 125(6):1342--1354, 1997.

\bibitem{pham2001stochastic}
Dinh~Tuan Pham.
\newblock Stochastic methods for sequential data assimilation in strongly
  nonlinear systems.
\newblock {\em Monthly Weather Review}, 129(5):1194--1207, 2001.

\bibitem{sakov2012iterative}
Pavel Sakov, Dean~S Oliver, and Laurent Bertino.
\newblock An iterative {EnKF} for strongly nonlinear systems.
\newblock {\em Monthly Weather Review}, 140(6):1988--2004, 2012.

\bibitem{sakov2008deterministic}
Pavel Sakov and Peter~R Oke.
\newblock A deterministic formulation of the ensemble {Kalman} filter: an
  alternative to ensemble square root filters.
\newblock {\em Tellus A: Dynamic Meteorology and Oceanography}, 60(2):361--371,
  2008.

\bibitem{arnold2013stochastic}
HM~Arnold, IM~Moroz, and TN~Palmer.
\newblock {Stochastic parametrizations and model uncertainty in the Lorenz’96
  system}.
\newblock {\em Philosophical Transactions of the Royal Society A: Mathematical,
  Physical and Engineering Sciences}, 371(1991):20110479, 2013.

\bibitem{maulik2019subgrid}
Romit Maulik, Omer San, Adil Rasheed, and Prakash Vedula.
\newblock Subgrid modelling for two-dimensional turbulence using neural
  networks.
\newblock {\em Journal of Fluid Mechanics}, 858:122--144, 2019.

\bibitem{vlachas2020backpropagation}
Pantelis~R Vlachas, Jaideep Pathak, Brian~R Hunt, Themistoklis~P Sapsis,
  Michelle Girvan, Edward Ott, and Petros Koumoutsakos.
\newblock Backpropagation algorithms and reservoir computing in recurrent
  neural networks for the forecasting of complex spatiotemporal dynamics.
\newblock {\em Neural Networks}, 2020.

\bibitem{maulik2020non}
Romit Maulik, Bethany Lusch, and Prasanna Balaprakash.
\newblock Non-autoregressive time-series methods for stable parametric
  reduced-order models.
\newblock {\em Physics of Fluids}, 32(8):087115, 2020.

\bibitem{hochreiter1997long}
Sepp Hochreiter and J{\"u}rgen Schmidhuber.
\newblock Long short-term memory.
\newblock {\em Neural Computation}, 9(8):1735--1780, 1997.

\bibitem{werbos1990backpropagation}
Paul~J Werbos.
\newblock Backpropagation through time: what it does and how to do it.
\newblock {\em Proceedings of the IEEE}, 78(10):1550--1560, 1990.

\bibitem{takens1981detecting}
Floris Takens.
\newblock Detecting strange attractors in turbulence.
\newblock In {\em Dynamical systems and turbulence, Warwick 1980}, pages
  366--381. Springer, 1981.

\bibitem{cechin2008optimizing}
Adelmo~L Cechin, Denise~R Pechmann, and Luiz~PL de~Oliveira.
\newblock {Optimizing Markovian modeling of chaotic systems with recurrent
  neural networks}.
\newblock {\em Chaos, Solitons \& Fractals}, 37(5):1317--1327, 2008.

\bibitem{dubois2020data}
Pierre Dubois, Thomas Gomez, Laurent Planckaert, and Laurent Perret.
\newblock {Data-driven predictions of the Lorenz system}.
\newblock {\em Physica D: Nonlinear Phenomena}, 408:132495, 2020.

\bibitem{gers2002applying}
Felix~A Gers, Douglas Eck, and J{\"u}rgen Schmidhuber.
\newblock {Applying LSTM to time series predictable through time-window
  approaches}.
\newblock In {\em Neural Nets WIRN Vietri-01}, pages 193--200. Springer, 2002.

\bibitem{bengio2015scheduled}
Samy Bengio, Oriol Vinyals, Navdeep Jaitly, and Noam Shazeer.
\newblock Scheduled sampling for sequence prediction with recurrent neural
  networks.
\newblock {\em arXiv preprint arXiv:1506.03099}, 2015.

\bibitem{sangiorgio2020robustness}
Matteo Sangiorgio and Fabio Dercole.
\newblock {Robustness of LSTM neural networks for multi-step forecasting of
  chaotic time series}.
\newblock {\em Chaos, Solitons \& Fractals}, 139:110045, 2020.

\bibitem{navon2009data}
Ionel~M Navon.
\newblock Data assimilation for numerical weather prediction: a review.
\newblock In {\em Data assimilation for atmospheric, oceanic and hydrologic
  applications}, pages 21--65. Springer, 2009.

\bibitem{lewis2006dynamic}
John~M Lewis, Sivaramakrishnan Lakshmivarahan, and Sudarshan Dhall.
\newblock {\em Dynamic data assimilation: a least squares approach}, volume
  104.
\newblock Cambridge University Press, Cambridge, 2006.

\bibitem{simon2006optimal}
Dan Simon.
\newblock {\em {Optimal state estimation: {Kalman}, H infinity, and nonlinear
  approaches}}.
\newblock John Wiley \& Sons, New York, 2006.

\bibitem{evensen2009data}
Geir Evensen.
\newblock {\em {Data assimilation: the ensemble {Kalman} filter}}.
\newblock Springer Science \& Business Media, 2009.

\bibitem{burgers1998analysis}
Gerrit Burgers, Peter Jan~van Leeuwen, and Geir Evensen.
\newblock Analysis scheme in the ensemble {Kalman} filter.
\newblock {\em {Monthly Weather Review}}, 126(6):1719--1724, 1998.

\bibitem{kalman1960new}
Rudolph~Emil Kalman.
\newblock A new approach to linear filtering and prediction problems.
\newblock 1960.

\bibitem{houtekamer2005ensemble}
Peter~L Houtekamer and Herschel~L Mitchell.
\newblock Ensemble {K}alman filtering.
\newblock {\em Quarterly Journal of the Royal Meteorological Society: A journal
  of the atmospheric sciences, applied meteorology and physical oceanography},
  131(613):3269--3289, 2005.

\bibitem{raanes2019adaptive}
Patrick~N. Raanes, Marc Bocquet, and Alberto Carrassi.
\newblock {Adaptive covariance inflation in the ensemble Kalman filter by
  Gaussian scale mixtures}.
\newblock {\em Quarterly Journal of the Royal Meteorological Society},
  145(718):53--75, 2019.

\bibitem{attia2019dates}
Ahmed Attia and Adrian Sandu.
\newblock Dates: a highly extensible data assimilation testing suite v1. 0.
\newblock {\em Geoscientific Model Development}, 12(2):629--649, 2019.

\bibitem{ahmed2020pyda}
Shady~E Ahmed, Suraj Pawar, and Omer San.
\newblock {PyDA: A Hands-On Introduction to Dynamical Data Assimilation with
  Python}.
\newblock {\em Fluids}, 5(4):225, 2020.

\bibitem{anderson1999monte}
Jeffrey~L Anderson and Stephen~L Anderson.
\newblock A {Monte Carlo} implementation of the nonlinear filtering problem to
  produce ensemble assimilations and forecasts.
\newblock {\em Monthly Weather Review}, 127(12):2741--2758, 1999.

\bibitem{grudzien2018chaotic}
Colin Grudzien, Alberto Carrassi, and Marc Bocquet.
\newblock {Chaotic dynamics and the role of covariance inflation for reduced
  rank Kalman filters with model error}.
\newblock {\em Nonlinear Processes in Geophysics}, 25(3):633--648, 2018.

\bibitem{hamill2001distance}
Thomas~M Hamill, Jeffrey~S Whitaker, and Chris Snyder.
\newblock Distance-dependent filtering of background error covariance estimates
  in an ensemble {Kalman} filter.
\newblock {\em Monthly Weather Review}, 129(11):2776--2790, 2001.

\bibitem{kepert2009covariance}
Jeffrey~D Kepert.
\newblock {Covariance localisation and balance in an ensemble Kalman filter}.
\newblock {\em Quarterly Journal of the Royal Meteorological Society},
  135(642):1157--1176, 2009.

\bibitem{houtekamer1998data}
Peter~L Houtekamer and Herschel~L Mitchell.
\newblock Data assimilation using an ensemble {Kalman} filter technique.
\newblock {\em Monthly Weather Review}, 126(3):796--811, 1998.

\bibitem{whitaker2012evaluating}
Jeffrey~S Whitaker and Thomas~M Hamill.
\newblock Evaluating methods to account for system errors in ensemble data
  assimilation.
\newblock {\em Monthly Weather Review}, 140(9):3078--3089, 2012.

\bibitem{anderson2007adaptive}
Jeffrey~L Anderson.
\newblock An adaptive covariance inflation error correction algorithm for
  ensemble filters.
\newblock {\em Tellus A: Dynamic Meteorology and Oceanography}, 59(2):210--224,
  2007.

\bibitem{kirchgessner2014choice}
Paul Kirchgessner, Lars Nerger, and Angelika Bunse-Gerstner.
\newblock {On the choice of an optimal localization radius in ensemble Kalman
  filter methods}.
\newblock {\em Monthly Weather Review}, 142(6):2165--2175, 2014.

\bibitem{attia2018optimal}
Ahmed Attia and Emil Constantinescu.
\newblock An optimal experimental design framework for adaptive inflation and
  covariance localization for ensemble filters.
\newblock {\em arXiv preprint arXiv:1806.10655}, 2018.

\end{thebibliography}

\end{document}